\documentclass[usenatbib,fleqn]{mnras}

\usepackage{mathptmx} % Change font to more like MNRAS
\usepackage{amsmath} % Additional maths symbols
\usepackage{graphicx} % Allows the use of adding images, or something like that
\usepackage{hyperref} % This sometimes goes mad if an equation is broken over a page and pdf latex exits with some unintelligible error
\usepackage{xspace} % gives option \xspace which is good for choosing intelligent spacing after newly defined commands
\usepackage{xcolor} % Enables coloured text
\usepackage{upgreek} % For upright greek characters
\usepackage[caption = false]{subfig} % Can have many plots on the same figure

\newcommand{\bold}[1]{\mathbf{#1}}
\newcommand{\dotp}[2]{\mathbf{#1}\cdot\mathbf{#2}}

\newcommand{\ft}[3]{\int{#1}e^{-i\dotp{#2}{#3}}\,\mathrm{d}^3{#3}}

\newcommand{\average}[1]{\langle#1\rangle}

\renewcommand{\pi}{\uppi}

\newcommand{\Dirac}[1]{\delta_\mathrm{D}(#1)}
\newcommand{\diff}{\mathrm{d}}

\newcommand{\Mpc}{\,h^{-1}\mathrm{Mpc}}

\newcommand{\iMpc}{\,h\mathrm{Mpc}^{-1}}

\newcommand{\Msun}{\,h^{-1}\mathrm{M_\odot}}

\newcommand{\Om}{\Omega_\mathrm{m}}
\newcommand{\Ov}{\Omega_\mathrm{v}}

\newcommand{\Ob}{\Omega_\mathrm{b}}

\newcommand{\Bnl}{\beta^\mathrm{NL}}

\newcommand{\eg}{e.g.,\xspace}

\newcommand{\nbody}{$N$-body\xspace}
\newcommand{\LCDM}{$\Lambda$CDM\xspace}
\newcommand{\halofit}{\textsc{halofit}\xspace}
\newcommand{\darkquest}{\textsc{dark quest}\xspace}
\newcommand{\wmap}{\textit{WMAP}\xspace}
\newcommand{\hmcode}{\textsc{hmcode}\xspace}
\newcommand{\camb}{\textsc{camb}\xspace}
\newcommand{\multidark}{\textsc{multidark}\xspace}
\newcommand{\bolshoi}{\textsc{bolshoi}\xspace}
\newcommand{\rockstar}{\textsc{rockstar}\xspace}

\newcommand{\web}[1]{\href{#1}{#1}}

\newcommand{\code}{\web{https://github.com/alexander-mead/BNL}\xspace}
\newcommand{\cosmosimaddress}{\web{https://www.cosmosim.org}\xspace}

\def\mnras{MNRAS}
\def\apj{ApJ}
\def\apjl{ApJ Let.}
\def\prd{Phys. Rev. D}
\def\aap{A\&A}
\def\jcap{J. Cosmol. Astropart. Phys.}
\def\apjs{ApJS}

\def\gtsim{\mathrel{\lower0.6ex\hbox{$\buildrel {\textstyle >}\over {\scriptstyle \sim}$}}}
\def\ltsim{\mathrel{\lower0.6ex\hbox{$\buildrel {\textstyle <}\over {\scriptstyle \sim}$}}}

\title[Beyond linear halo bias]{Including beyond-linear halo bias in halo models}
\author[A. J. Mead et al.]{A. J. Mead$^{1, 2, 3}$\thanks{E-mail: alexander.j.mead@googlemail.com}
and
L. Verde$^{2, 4}$
 \\
1 Institute for Astronomy, University of Edinburgh, Royal Observatory, Blackford Hill, Edinburgh, EH9 3HJ, UK\\
2 Institut de Ci\`encies del Cosmos, Universitat de Barcelona, Mart\'i Franqu\`es 1, Barcelona 08028, Spain\\
3 Department of Physics and Astronomy, University of British Columbia, 6224 Agricultural Road, Vancouver, BC V6T 1Z1, Canada\\
4 ICREA, Instituci\'o Catalana de Recerca i Estudis Avan\c{c}ats, Passeig Llu\'is Companys 23, Barcelona 08010, Spain
}

\date{Accepted 2021 March 9. Received 2021 February 10; in original form 2020 November 17}
\pagerange{\pageref{firstpage}--\pageref{lastpage}} 
\pubyear{2021}

\begin{document}
\maketitle

\label{firstpage}

% Abstract
\begin{abstract}
We derive a simple prescription for including beyond-linear halo bias within the standard, analytical halo-model power spectrum calculation. This results in a corrective term that is added to the usual two-halo term. We measure this correction using data from \nbody simulations and demonstrate that it can boost power in the two-halo term by a factor of $\sim2$ at scales $k\sim0.7\iMpc$, with the exact magnitude of the boost determined by the specific pair of fields in the two-point function. How this translates to the full power spectrum depends on the relative strength of the one-halo term, which can mask the importance of this correction to a greater or lesser degree, again depending on the fields. Generally we find that our correction is more important for signals that arise from lower-mass haloes. When comparing our calculation to simulated data we find that the under-prediction of power in the transition region between the two- and one-halo terms, which typically plagues halo-model calculations, is almost completely eliminated when including the full non-linear halo bias. We show improved results for the auto and cross spectra of galaxies, haloes and matter. In the specific case of matter--matter or matter--halo power we note that a large fraction of the improvement comes from the non-linear biasing between low- and high-mass haloes. We envisage our model being useful in the analytical modelling of cross correlation signals. Our non-linear bias halo-model code is available at \code.
\end{abstract}

%We derive a simple prescription for including the non-linearity in halo biasing within the standard, analytical halo-model power spectrum calculation. 

\begin{keywords}
cosmology: theory -- 
large-scale structure of Universe
%\vspace{0.5cm}
\end{keywords}

\section{Introduction}
\label{sec:introduction}

% Halo model successes
The halo model \citep[reviewed by][]{Cooray2002} is widely used in the interpretation of data from cosmological large-scale structure surveys. The model is not derived from first principles, and is instead phenomenological in that it is a description of the properties of the universe as seen in \nbody simulations. In the model, all matter is taken to reside in haloes that trace the large-scale matter fluctuations in a biased way, with this `halo bias' usually taken to be linear with respect to the underlying linear matter field. The model also makes a number of other simplifying assumptions; usually that haloes are spherical, devoid of substructure, and that the halo mass determines all of the halo properties with no scatter. In addition, choices must be made for the mass function, biasing recipe, and profile for haloes. Fields, be they sourced by point tracers or else via some emissivity, can then be assumed to occupy haloes in different ways depending on the halo properties; the model will then make predictions for the $n$-point correlation functions. The halo model has been successful in explaining the broad shape of the galaxy--galaxy correlation function \citep[\eg][]{Seljak2000, Peacock2000}, as well as making inroads in the understanding of the connection between galaxy formation and the haloes in which the galaxies reside \citep[\eg][]{Mandelbaum2005, Cacciato2012} and is also widely used to describe other cross correlations \citep[\eg][]{Addison2012, Hill2014, Ma2015, Padmanabhan2017b, Hill2018, Wolz2019, Tanimura2019b, Koukoufilippas2020}.

% Halo model problems
It has been $20$ years since the modern formulation of the halo model came to prominence, and since then the quality of data to which the halo model is exposed have increased dramatically. With this in mind, it is reasonable to revisit some of the foundational assumptions. The two-point model breaks the clustering signal down in to a sum of two parts: a `two-halo' term that originates from the clustering between different pairs of haloes, and a `one-halo' term that  originates from clustering within the same halo. For all cosmological observables the two-halo component of the clustering dominates the signal at large scales while the one-halo component dominates at small scales. One perennial problem with the model has been the `transition' region, where both two- and one-halo terms have similar magnitude and both contribute to the predicted signal. In general the model underpredicts the strength of clustering in this region, for example \cite{Fedeli2014b} and \cite{Mead2015b} showed that this underprediction can be as much as $30$ per cent for the matter--matter power spectrum, with the exact amount depending on redshift and cosmology. \cite{Tinker2005} also note similar problems in halo-model descriptions of the transition region of the galaxy--galaxy correlation function. This region of the spectrum is also called the `quasi-linear' regime because the evolution of perturbations at these scales is not exactly governed by linear perturbation theory, but there is hope that it can be understood via higher-order perturbative schemes. The fact that the transition and quasi-linear regions coincide is no coincidence as it is inevitable that halo formation is linked to scales where perturbative descriptions break down.

% Solutions for matter-matter
In the special case of the matter--matter power spectrum, the inaccuracies of the halo model are `remedied' by devising fitting functions (\eg \halofit of \citealt{Smith2003, Takahashi2012}) or else by adding phenomenological parameters (\eg \hmcode \citealt{Mead2015b, Mead2016, Mead2021a}) and fitting these to power spectra measured from high-resolution cosmological \nbody simulations. In this way, models of the power spectrum that are accurate at the $5$ per-cent level for $z<2$ and $k<10\iMpc$ have been developed. Alternatively, \cite{Valageas2011}, \cite{Mohammed2014a}, \cite{Seljak2015} and \cite{Philcox2020} have all improved predictions at quasi-linear scales by incorporating perturbation theory for the matter field, with per-cent level accuracy for $k<1\iMpc$ reported for some models. However, it could be argued that predictions of the matter--matter power spectrum are one of the least useful applications of the halo model, given that this spectrum can be accurately measured from \nbody simulations (at least, if one ignores the inconvenient issue of baryonic feedback). Indeed, the most accurate predictors for the matter--matter spectrum as a function of cosmology for $k<10\iMpc$ come from `emulation' schemes \citep[\eg][]{Lawrence2010, Agarwal2012, Agarwal2014, Lawrence2017, Knabenhans2019} in which measured power spectra from simulations that span a range of cosmologies are interpolated between.

% The point of this paper
The models discussed for the matter--matter power in the previous paragraph are difficult to generalise to spectra other than matter--matter. For example, if one is interested in the halo-model prediction for anything connected with galaxies, including the connection between matter and galaxies, then one is reduced to using the standard halo model, which comes with the standard problems. In this paper, we are interested in improving the halo-model predictions around the transition region for any field pair, and we do not restrict our focus to the matter--matter spectrum. We focus our attention on the non-linear portion of the halo bias, a treatment of which is almost always absent from standard halo-model calculations. We aim to do this while doing minimal damage to the existing halo-model apparatus, because this is widely used by the community in its standard form. 

% Layout
In Section~\ref{sec:halo_model} we present the standard derivation of the halo-model power spectrum and we demonstrate how to include non-linear halo bias in the calculation in a clean, isolated way. In Section~\ref{sec:measurement} we show how we calculate the new non-linear halo bias term, which involves measurements of this new term from \nbody simulations. In Section~\ref{sec:results} we present the results of including this new term in halo-model calculations for matter--matter, halo--matter and halo--halo power spectra and we show an application to power spectra involving galaxies. Finally, we conclude in Section~\ref{sec:summary} and propose some ideas for future work. Appendix~\ref{app:numerical_calculations} discusses the technical details of how we incorporate the new non-linear bias term in our numerical calculations. Appendix~\ref{app:additional_redshifts} presents results at redshifts other than those presented in the main body of the paper. Appendix~\ref{app:exact_biasing} presents a pedagogical example mock universe where linear biasing is exactly respected, and we demonstrate that this can be modelled essentially perfectly by the standard halo-model calculation.

\section{Halo model}
\label{sec:halo_model}

\subsection{Standard derivation}

% Introduction
We shall first present the standard derivation of the halo model power spectrum (\eg \citealt{Cooray2002}). This subsection can be skipped by those readers familiar with the halo model, but is included for completeness. We carry out this derivation in a comoving, periodic volume $V$ because we believe this is the most transparent. Note that this means that the Fourier modes are discrete. We eventually take $V\to\infty$ to retrieve the continuum limit. To keep the notation simple we suppress any time dependence in function arguments, but the reader should recall that most functions can be time as well as (Fourier) space dependent.

% Standard derivation of two- and one-halo terms
We consider the power spectrum between a pair of three-dimensional cosmological `fields', that could be identical. Examples of such fields would be `matter', `halo' or `galaxy' over-densities that vary from place-to-place in the Universe, and that could require smoothing to be reasonably defined. These fields could be further restricted to be haloes in a specific mass range or galaxies of a specific type. They could also be fields that are generated by some emissive process, such as infrared light or electron pressure. We define fields $\theta_u(\bold{x})$, where  $\bold{x}$ is comoving position and the label $u$ stands for the field. We will be interested in power spectra between pairs of such fields, $P_{uv}(k)$, which has units of the product of the units of fields $u$ and $v$ and an additional factor of volume.

% Derivation
We start from the assumption that all fields reside in haloes, which allows us to write the total field at some position as a sum of contributions, $\theta_{u,i}$, from haloes with positions $\bold{x}_i$
\begin{equation}
\theta_u(\mathbf{x})=\sum_i \theta_{u,i}(\mathbf{x}-\mathbf{x}_i)\ ,
\label{eq:theta_haloes}
\end{equation}
if we allow our definition of `halo' to be sufficiently general then this is always true: we can always break down the total field into a sum of contributions. For example, $\theta_{u,i}$ could be the `halo profile' for matter, but for other fields it could be an emission profile. In what follows we refer to $\theta_{u,i}$ as the halo profile, but one should keep in mind that this is a more general quantity. We can write the Fourier transform, in terms of comoving wavenumber $\bold{k}$, as
\begin{equation}
\theta_{u,\bold{k}}=\frac{1}{V}\ft{\theta_u(\bold{x})}{\bold{k}}{\bold{x}}=
\frac{1}{V}\sum_i\ft{\theta_{u,i}(\mathbf{x}-\mathbf{x}_i)}{\bold{k}}{\bold{x}}\ ,
\end{equation}
where the factor of $1/V$ ensures that the units of $\theta_{u,\bold{k}}$ are the same of those of $\theta_u(\bold{x})$. If we make the variable change $\bold{y}=\bold{x}-\bold{x}_i$ then we can write
\begin{equation}
\theta_{u,\bold{k}}=\frac{1}{V}\sum_i\mathrm{e}^{-i\dotp{k}{x_i}}\ft{\theta_{u,i}(\bold{y})}{\bold{k}}{\bold{y}}=
\frac{1}{V}\sum_i\mathrm{e}^{-i\dotp{k}{x_i}}W_{u,i}(k)\ ,
\end{equation}
where we have defined
\begin{equation}
W_{u,i}(k)=\int_0^\infty\frac{\sin(kr)}{kr}\theta_{u,i}(r)4\pi r^2\,\diff r\ ,
\label{eq:window_function}
\end{equation}
as a Fourier transform of the halo profile, which has units of field $u$ multiplied by volume. We have made the additional assumption that the halo profile is spherically symmetric, which allows us to write $W_{u,i}$ in terms of $k=|\bold k|$ and $r=|\bold r|$ and also ensures that $W_{u,i}$ is real. Let us now construct the power spectrum between two fields (which could be identical), $\theta_u$ and $\theta_v$:
\begin{equation}
P_{uv}(k)=\average{\theta^{*}_{u,\bold{k}}\theta^{}_{v,\bold{k}}}=\average{\frac{1}{V^2}\sum_i\sum_j \mathrm{e}^{-i\dotp{k}{(x_i-x_j)}}W_{u,i}(k)W_{v,j}(k)}\ .
\end{equation}
We can break this equation up into two pieces, the $i=j$ piece, which we call the one-halo term and the $i\neq j$ piece, which we call the two-halo term. The one-halo term is
\begin{equation}
P^\mathrm{1H}_{uv}(k)=\average{\frac{1}{V^2}\sum_i W_{u,i}(k)W_{v,i}(k)}\ ,
\end{equation}
where the expectation value is taken over all modes in the volume and we have made the standard assumptions about isotropy and homogeneity that ensure that we can write expressions as a function of $k$ only. We now assume a continuum of haloes, labelled with a mass $M$, and that the halo mass is the sole determinant of the halo properties. We further assume that these haloes are distributed according to a mass-distribution function $n(M)$ (sometimes denoted $\diff n/\diff M$ in the literature), where $n(M)\,\diff M$ describes the number-density of haloes in the range $M$ to $M+\diff M$. We can then write the sum as an integral  by taking $V\to\infty$ using
\begin{equation}
\frac{1}{V}\sum_i \to \int n(M)\,\diff M\ ,
\end{equation}
to get
\begin{equation}
P^\mathrm{1H}_{uv}(k)=\int_0^\infty W_{u}(M,k)W_{v}(M,k) n(M)\,\diff M\ .
\label{eq:one_halo_term}
\end{equation}
We can apply the same reasoning to the two-halo term, where we convert both sums to integrals
\begin{multline}
P^\mathrm{2H}_{uv}(k)=\int_0^\infty\int_0^\infty P_\mathrm{hh}(M_1,M_2,k)\times \\
W_{u}(M_1,k)W_{v}(M_2,k)n(M_1)n(M_2)\,\diff M_1\diff M_2\ ,
\label{eq:full_two_halo_term}
\end{multline}
and we recognise the expectation of the complex exponential to be the power spectrum of the halo centres: $P_\mathrm{hh}(M_1,M_2,k)$. Note that the functions $W_u$ and $W_v$ are common between equations~(\ref{eq:one_halo_term}) and  (\ref{eq:full_two_halo_term}). At this stage of the derivation it is common to make the approximation that haloes are linearly biased tracers of the underlying linear matter field
\begin{equation}
P_\mathrm{hh}(M_1,M_2,k)\simeq b(M_1)b(M_2)P^\mathrm{lin}_\mathrm{mm}(k)\ ,
\label{eq:linear_bias}
\end{equation}
where $b(M)$ is the linear halo bias of haloes with mass $M$ and $P^\mathrm{lin}_\mathrm{mm}(k)$ is the linear-theory matter power spectrum. This approximation has the virtue of being correct at very large scales and also is mathematically convenient because it allows us to split the double integral in equation~(\ref{eq:full_two_halo_term}) into a product of two one-dimensional integrals of similar form. The final result is then
\begin{equation}
P^{2\mathrm{H}}_{uv}(k)=P^\mathrm{lin}_\mathrm{mm}(k)
\prod_{n=u,v}\left[\int_0^\infty W_n(M,k)b(M)n(M)\,\mathrm{d}M\right]\ .
\label{eq:standard_two_halo_term}
\end{equation}

% Normalisation of mass function and bias
The adopted halo mass function and linear halo bias \emph{must} satisfy the following properties for any power spectrum involving the matter to have the correct large-scale limit\footnote{Achieving these limits is difficult numerically because of the large amount of mass contained in low-mass haloes according to most popular mass functions. Special care must be taken with the two-halo integral in the case of power spectra that involve the matter field. See Appendix \ref{app:numerical_calculations}}:
\begin{equation}
\int_0^\infty Mn(M)\,\mathrm{d}M=\bar\rho\ ,
\label{eq:mass_normalisation}
\end{equation}
\begin{equation}
\int_0^\infty Mb(M)n(M)\,\mathrm{d}M=\bar\rho\ .
\label{eq:bias_normalisation}
\end{equation}
where $\bar\rho$ is the mean comoving cosmological matter density. In words, these equations enforce that all matter is contained in haloes and that, on average, matter is unbiased with respect to itself. 

% Example for matter P(k) (maybe remove)
As an example of how the two-halo term works in practice: In the special case of the matter--matter overdensity power spectrum we can write $\theta_\mathrm{m}(M,r) = \rho_\mathrm{m}(M,r)/\bar\rho$ where $\rho_\mathrm{m}(M,r)$ is the halo matter density profile. We then note that $W_\mathrm{m}(M,k\to0)=M/\bar\rho$, and it is usual to factorise this normalisation, such that $W_\mathrm{m}(M,k)=MU_\mathrm{m}(M,k)/\bar\rho$, with $U_\mathrm{m}(M, k\to0)=1$. We can then write equation~(\ref{eq:standard_two_halo_term}) as 
\begin{equation}
P^{2\mathrm{H}}_{uv}(k)=P^\mathrm{lin}_\mathrm{mm}(k)
\left[\frac{1}{\bar\rho}\int_0^\infty M U_\mathrm{m}(M,k)b(M)n(M)\,\mathrm{d}M\right]^2\ .
\label{eq:standard_two_halo_term_matter}
\end{equation}
and we see that $P^{2\mathrm{H}}_{uv}(k\to0)=P^\mathrm{lin}_\mathrm{mm}(k\to0)$ automatically as the term in the square brackets equals unity in this limit. For spectra other than matter--matter this is no longer true, and in general the large-scale limit of the two halo term will be equal to the linear spectrum multiplied by amplitude factors that account for the field content and the bias that arises from how these fields populate haloes. 

\subsection{Including non-linear halo bias}

% Approximations
It is worth examining some of the approximations that lead to the standard halo-model equations~(\ref{eq:one_halo_term}) and (\ref{eq:standard_two_halo_term}): It has been assumed that halo profiles are perfectly spherical with no substructure, that there is no scatter in profile properties at fixed host halo mass, and that halo properties depend only on halo mass and not on other properties, for example halo location. These approximations will break down, and the errors in the eventual power spectrum that they contribute will vary with the fields that are being considered \citep[\eg][]{Smith2005, Giocoli2010, Smith2011b, Chen2020, Voivodic2020}. It is also usually assumed that haloes trace the underlying linear matter distribution with a linear halo bias. In this paper we focus on including scale-dependent, non-linear halo bias within the halo-model calculation. A previous attempt to include non-linear halo bias has been made by \cite*{Smith2007} where the combination of standard perturbation theory (SPT) and an Eulerian bias expansion were used to model the matter--matter, matter--halo and halo--halo power spectra. This approach was demonstrated to be successful at very large scales, where perturbation theory is a good description of clustering. A similar model is presented by \cite{McDonald2006b} where the renormalisation of coefficients in the bias expansion was considered for the first time. Both of these models rely on perturbation theory, and they fail on smaller scales where perturbation theory breaks down and where much of the constraining power of a contemporary cosmological survey lies. Alternatively, \cite{Fedeli2014b} investigate a phenomenological non-linear biasing model where $b(M)\to b(M,k)$ in equation~(\ref{eq:standard_two_halo_term}) and the $k$ dependence is fitted to \nbody data. In this paper we follow a different approach, and use measurements of the non-linear halo bias from simulations to push more deeply in to the non-linear regime.

% Galaxy-galaxy lensing
Our study is of particular relevance to galaxy--galaxy lensing, which is the study of the two-point function between galaxy and matter overdensities, where the matter clustering is accessed via weak gravitational lensing. The halo model is very widely used in the interpretation of data from such observations, and is used to understand the fundamental link between haloes and galaxy formation therein \citep[\eg][]{Mandelbaum2005, Cacciato2009}. Most halo models of galaxy--galaxy lensing assume a linear halo bias \citep[\eg][]{Cacciato2012, Dvornik2018}, but the halo model of the galaxy--galaxy lensing signal developed by \cite{vandenBosch2013} includes some non-linear effects of halo biasing in two ways. First, the idea that haloes cannot overlap (so-called halo exclusion) is considered by forcing the halo--halo correlation function to $-1$ on scales below the sum of the virial radii of the haloes contributing to the two-point function. Second, scale-dependent halo bias is included using a fitting function for the `radial-dependent' bias taken from \cite{Tinker2005}. This has the advantage that the \cite{Tinker2005} result is returned automatically on quasi-liner scales, but the disadvantage that the applicability of results is limited to a specific galaxy population. In the \cite{vandenBosch2013} model, and in some other galaxy--galaxy lensing prescriptions, the halo bias is defined to be relative to the non-linear matter field. In our work we avoid this, since in principle a good halo model should be able to predict the non-linear power spectra of matter--matter, matter--galaxies and galaxies--galaxies all from the same set of founding assumptions. 

% Transition region
Different assumptions regarding non-linear halo bias impact on halo-model predictions in the transition region between the two- and one-halo terms, an accurate modelling of which is becoming important given the quality of contemporary data; traditional halo models generally underpredict either power or correlation in the transition region. In \cite{Hayashi2008} the halo--matter cross correlation is modelled by adopting the maximum of either the two- or one-halo term in the transition region. In \cite{Garcia2020} it is demonstrated that the transition region can be well modelled if halo boundaries are tailored to the cross correlation and if a halo exclusion term is included. In the \cite{Koukoufilippas2020} measurement of the the thermal Sunyaev-Zel'dovich (tSZ)--galaxy cross correlation the transition region was scaled by the ratio of \halofit to the standard halo model prediction for the matter--matter power spectrum. In \cite{Hang2021} the galaxy auto correlation is assumed to be a constant bias multiplying a linear part, plus a different constant bias multiplying a non-linear part, which is taken from \halofit. None of these solutions are particularly appealing from a theoretical perspective. We note that it is not obvious that the same correction that works in \halofit for matter--matter would apply more generally, and indeed later we show that this is not the case.

% Discussion of the standard two-halo term
In the standard halo-model derivation, the calculation of the two-halo term is made tractable by making the approximation that the haloes are \emph{linearly} biased tracers of the underlying \emph{linear} matter field (equation~\ref{eq:linear_bias}). At large scales the shape of the two-halo term of any power spectrum computed in this way will be exactly that of the linear spectrum; scale-dependent deviations from this arise due to the factors of $W_n(M,k)$ in equation~(\ref{eq:standard_two_halo_term}). This means that the shape of the standard two-halo term is only different from linear theory on scales corresponding to the virial radii of the most massive haloes where $W_n(M,k)$ starts to deviate from constant. We note that it seems unlikely that a linear halo bias is a good description of the clustering relation between haloes and matter on scales comparable to the sizes of individual haloes.

%This is numerically convenient because it allows us to bring $P^\mathrm{lin}_\mathrm{mm}(k)$ out of the integrals for the two-halo term and leaves two integrals that are independent from each other, thus reducing a two-dimensional integral to the product of two one-dimensional integrals. This also ensures that the halo-model power spectrum that we compute has the correct large-scale limit. 

% Including non-linear halo bias measured from simulations
Given that we know that in reality halo bias is not linear, let us instead not make this approximation, but write the halo power spectrum in the following way
\begin{equation}
P_\mathrm{hh}(M_1,M_2,k)=b(M_1)b(M_2)P^\mathrm{lin}_\mathrm{mm}(k)[1+\Bnl(M_1,M_2,k)]\ ,
\label{eq:Bnl_def}
\end{equation}
where the function $\Bnl$ captures all the things missing from the standard linear-bias--linear-field model. We know some things about $\Bnl$, specifically the large-scale limit: $\Bnl(M_1,M_2,k\to0)=0$, and also that it must obey symmetry with respect to mass arguments: $\Bnl(M_1,M_2,k)=\Bnl(M_2,M_1,k)$. The new idea presented in this paper is to include $\Bnl$ within semi-analytical calculations using the halo model. If we substitute equation~(\ref{eq:Bnl_def}) into equation~(\ref{eq:full_two_halo_term}) we have
\begin{multline}
P^{2\mathrm{H}}_{uv}(k)=P^\mathrm{lin}_\mathrm{mm}(k)\prod_{n=u,v}\left[\int_0^\infty W_n(M,k)b(M)n(M)\mathrm{d}M\right] + \\
P^\mathrm{lin}_\mathrm{mm}(k)I^\mathrm{NL}_{uv}(k)\ .
\label{eq:new_two_halo_term}
\end{multline}
The first term is standard, and is identical to equation~(\ref{eq:standard_two_halo_term}) and the new content is captured by $I^\mathrm{NL}_{uv}(k)$ in the second term
\begin{multline}
I^{\mathrm{NL}}_{uv}(k)=\int_0^\infty\int_0^\infty \Bnl(M_1,M_2,k)\times \\
W_u(M_1,k)W_v(M_2,k)b(M_1)b(M_2)n(M_1)n(M_2)\,\mathrm{d}M_1\mathrm{d}M_2\ .
\label{eq:Inl}
\end{multline}
It is worth considering what this new content represents. The standard approximation is that haloes are \emph{linearly} biased tracers of an underlying \emph{linear} matter field. Since we know the linear matter field to be a Gaussian random field this implies that the halo fields themselves must be Gaussian random. The new content is all departures from this simple picture, including enhanced halo clustering, filamentary, sheet and void structure; all of which can be realised by moving haloes from their linear--Gaussian locations. This is often termed `scale-dependent' bias. There is no reason to assume that the function $\Bnl$ should be particularly simple, indeed, given the complexity of the new content we might expect it to be a complicated function. We also have no reason to believe that it should be independent of either redshift or cosmological parameters. Since we know perturbation theory to be a good description of the Universe for $k\ltsim0.2\iMpc$ the $\Bnl$ function must also contain these perturbative corrections if it is to be a good, general model of clustering.

\subsection{Relation to other non-linear bias definitions}

% Relation of new definition to other definitions
Non-linear halo bias is often discussed in terms of the scale-dependent function, $b_\mathrm{hh}(M_1,M_2,k)$, defined by
\begin{equation}
P_\mathrm{hh}(M_1,M_2,k)=b^2_\mathrm{hh}(M_1,M_2,k)P^\mathrm{lin}_\mathrm{mm}(k)\ ,
\label{eq:standard_auto_bias}
\end{equation}
where
\begin{equation}
b^2_\mathrm{hh}(M_1,M_2,k\to0)=b(M_1)b(M_2)\ .
\end{equation}
This is can be computed using our formalism (via equation~\ref{eq:Bnl_def}) as
\begin{equation}
b^2_\mathrm{hh}(M_1,M_2,k)=b(M_1)b(M_2)[1+\Bnl(M_1,M_2,k)]\ ,
\label{eq:auto_bias_conversion}
\end{equation}
if $M_1=M_2$ there is an additional one-halo contribution (in this case called the shot noise) of $1/\bar{n}_\mathrm{h}$ where $\bar{n}_\mathrm{h}$ is the mean number density of the halo sample. Some authors define a halo bias via the cross spectrum between the halo-number-density field and the matter overdensity field:
\begin{equation}
P_\mathrm{hm}(M,k)=b_\mathrm{hm}(M,k)P^\mathrm{lin}_\mathrm{mm}(k)\ .
\label{eq:standard_cross_bias}
\end{equation}
This halo bias will be generally different from that measured through the halo-number-density auto spectrum, although they coincide at large scales. This `cross bias' can be written in our notation (setting $u=\mathrm{h}$ and $v=\mathrm{m}$) as
\begin{multline}
b_\mathrm{hm}(M,k)=
%\frac{1}{n_\mathrm{h}(M)}\int_0^\infty W_\mathrm{m}(M_2,k)n(M_2)\,\mathrm{d}M_2+ \\
W_\mathrm{m}(M,k)/P^\mathrm{lin}_\mathrm{mm}(k)+\\
b(M)\int_0^\infty\mathrm{d}M_2[1+\Bnl(M,M_2,k)]W_\mathrm{m}(M_2,k)b(M_2)n(M_2)
\ .
\label{eq:cross_bias_conversion}
\end{multline}
The first term on the right-hand side of equation~(\ref{eq:cross_bias_conversion}) arises from the one-halo term (equation~\ref{eq:one_halo_term}) while the second is from the two-halo term (equation \ref{eq:new_two_halo_term}). The equations in this subsection can be derived by taking $W_\mathrm{h}(M',k)=\Dirac{M-M'}/\bar{n}_\mathrm{h}$ as the window function for the halo, therefore in both equations~(\ref{eq:auto_bias_conversion}) and (\ref{eq:cross_bias_conversion}) we have assumed that we are considering thin halo mass bins. 

% Halo bias with respect to the non-linear field.
Some authors define the halo biases in equation~(\ref{eq:standard_auto_bias}) or (\ref{eq:standard_cross_bias}) with respect to the \emph{non-linear} matter--matter power spectrum, rather than the linear matter spectrum. In this paper, we always define it with respect to the linear spectrum and this distinction is important in our work because: Firstly, it is halo bias defined in this way that enters standard halo-model calculations, and secondly a general model should be able to \emph{predict} the non-linear matter--matter spectrum, and since all matter is contained in haloes this itself \emph{must} come from the haloes. If one wanted to work with halo bias defined relative to the underlying \emph{non-linear} matter--matter spectrum then this itself is computable from our model by taking matter overdensity profiles in equations~(\ref{eq:new_two_halo_term}) and (\ref{eq:Inl}).

\section{Measuring non-linear halo bias}
\label{sec:measurement}

\subsection{Measurement}

% What exactly to measure
It would be ideal if we could calculate $\Bnl$ from first principles. However, while perturbation theory may be able to give us some insight at large scales, we anticipate $\Bnl$ to have structure on scales that are out of reach of even state-of-the-art perturbation theories. Therefore, we decide to measure the required function from \nbody simulations. If we simply rewrite equation~(\ref{eq:Bnl_def}) as 
\begin{equation}
1+\Bnl(M_1,M_2,k)=\frac{P_\mathrm{hh}(M_1,M_2,k)}{b(M_1)b(M_2)P^\mathrm{lin}_\mathrm{mm}(k)}\ ,
\label{eq:Bnl_rewrite}
\end{equation}
we can see that a sensible way to measure $\Bnl$ is to measure $P_\mathrm{hh}(M_1,M_2,k)$ and then to divide it by the linear power spectrum multiplied by the product of linear halo bias factors for $M_1$ and $M_2$. Recall that $P_\mathrm{hh}(M_1,M_2,k)$ is the cross-power spectrum measured between haloes of mass $M_1$ with those of mass $M_2$. In real measurements we have to bin haloes in mass to measure this function. If we consider the auto-spectrum ($M_1=M_2$) then we will also have to subtract (halo) shot noise from the measured halo power spectra, because, in our approach, this is the one-halo contribution to the halo auto spectrum, and we are interested only in the two-halo contribution. This shot noise is not a problem if we consider the cross spectrum between haloes in two different mass bins because it is automatically zero when the haloes in each leg of the cross spectrum are different. The shot-noise contribution to a given auto-halo-power-spectrum measurement is given by
\begin{equation}
P^\mathrm{SN}_\mathrm{hh}=\frac{1}{\bar n_\mathrm{h}}\ ,
\label{eq:shot_noise}
\end{equation}
where $\bar n_\mathrm{h}$ is the mean halo number density measured for the mass bin in question. Some authors consider a non-Poissonian shot noise, not given by equation~(\ref{eq:shot_noise}) and may also consider halo exclusion (the spatially exclusivity of haloes) to affect the shot-noise term. In our picture, halo exclusion will enter in the non-linear halo bias portion of the two-halo term, since it pertains to the way that haloes trace the underlying matter field, rather than the structure within the haloes themselves. In fact, regardless of the position one takes on a shot-noise correction for haloes, subtracting power according to equation~(\ref{eq:shot_noise}) is correct in our work given that this is exactly what we take for the one-halo term when we evaluate halo--halo auto power spectra using our model (equation~\ref{eq:one_halo_term}). Subtracting equation~(\ref{eq:shot_noise}) can therefore be seen as a method for isolating the two-halo term from the halo--halo auto power measurement.

\subsection{Simulations}

% Multidark and measuring the halo--halo power spectrum
We measure $\Bnl(M_1,M_2,k)$ using data taken from the \multidark simulation database\footnote{\cosmosimaddress} \citep{Klypin2011,Prada2012,Riebe2013}. We use data from the original \multidark simulation, which simulated $N=2048^3$ particles in a $L=1000\Mpc$ cube. We consider the combination of a high number of particles in the large volume of \multidark advantageous for our measurement at both large and small scales. We utilise haloes that have been identified via the \rockstar phase-space algorithm of \cite{Behroozi2013}\footnote{We obtain very similar results if we utilise haloes identified via the bound-density maximum (BDM) algorithm of \cite{Klypin1997}.} and defined using the `virial' criterion, with a spherical-overdensity threshold given by $\simeq 360\bar\rho$ at $z=0$, which comes from spherical-collapse calculations \citep[\eg][]{Bryan1998} for the simulated \wmap5 \LCDM cosmology: $\Om=0.27$; $\Ob=0.0469$; $\Ov=1-\Om$; $h=0.7$; $n=0.95$; $\sigma_8=0.82$. We prefer to use the virial halo definition of haloes because of results presented in \cite{Mead2021a}, where it was demonstrated that halo-model calculations are more robust with respect to cosmological dependence when haloes are defined using the virial condition (as opposed to $200\bar\rho$ or $200\rho_\mathrm{c}$). However, in principle one could work with any desired halo definition as long as consistency is maintained\footnote{This would necessitate remeasuring $\Bnl$ from a different halo catalogue with the new mass definition, or else translating between halo-mass definitions (\eg via the NFW profile).}. To measure $\Bnl$ (equation~\ref{eq:Bnl_rewrite}) we calculate halo--halo cross power between $8$ mass bins, which leads to $36$ unique cross-combinations. The choice of $8$ mass bins represents a reasonable compromise between having enough bins so that we gain in halo-mass resolution, while also allowing each bin to contain enough haloes such that the measurement is not too noisy\footnote{We obtain very similar results if we increase the number of mass bins from $8$ to $12$.}. The lower limit of our lowest-mass bin corresponds to haloes with $50$ particles, which in \multidark is $\simeq 10^{11.6}\Msun$. In this paper we only care about the halo position and mass being accurately measured, and $50$ particles can be considered a minimum for this \citep{Knebe2011}. We compared the mass function of our halo sample with theoretical expectations from \cite{Tinker2010} and find good agreement, even at our $50$ particle lower limit\footnote{We also use data from the \bolshoi simulation, which is the same as \multidark in all respects apart from being smaller, $250\Mpc$, and thus able to resolve lower-mass haloes.}.

% Discussion of nu vs. M
The limits of our mass bins are defined to be equally spaced in peak height, $\nu$, defined via
\begin{equation}
\nu=\frac{\delta_\mathrm{c}}{\sigma(M)}\ ,
\label{eq:nu}
\end{equation}
between the $\nu$ value corresponding to the lowest-mass haloes ($\nu\simeq0.74$ at $z=0$) and that corresponding to the most-massive halo ($\nu\simeq3.97$ at $z=0$). We prefer to work with $\nu$ bins, rather than $M$ or $\log(M)$, because many quantities of cosmological interest, such as the halo mass function, bias and concentration--mass relation, have been shown to be more independent of further cosmology dependence when expressed in terms of $\nu$. While we do not investigate the cosmology dependence of $\Bnl$ in this paper, we feel that expressing it in terms of $\nu$ may be useful in the future. In equation~(\ref{eq:nu}) we take the (cosmology-dependent) critical threshold for collapse, $\delta_\mathrm{c}$ from \cite{Nakamura1997}, although this cosmology dependence has a negligible impact (see \citealt{Mead2021a}) on the eventual results in this paper. The standard deviation in the linear matter field, $\sigma(M)$, is calculated analytically in the usual way using a top-hat filter. 

\subsection{Constructing $\Bnl$}

% Constructing B_NL from the halo-halo power, linear bias and linear matter power spectrum
We compute $P_\mathrm{hh}(M_1, M_2, k)$ via fast Fourier transform with $512^3$ cells and subtract halo shot noise for auto spectra as per equation~(\ref{eq:shot_noise}) in order to isolate the two-halo component. We only keep wavenumbers up to half of the Nyquist frequency (for our number of mesh cells $k_\mathrm{Ny}/2\simeq0.8\iMpc$) in order to avoid the effects of aliasing. Following equation~(\ref{eq:Bnl_rewrite}), we construct $\Bnl(M_1, M_2, k)$ by dividing the measured halo--halo power spectra by the linear power spectrum. For $k<0.08\iMpc$ we take the linear matter--matter spectrum to be that measured from the simulation itself\footnote{In fact, we take the $z\simeq 3$ density field grown to the desired redshift, because the initial conditions of these simulations no longer exist.}. At large scales this allows us to cancel some cosmic variance that would otherwise dominate our measurement. We are unable to use the simulation measurement for $k>0.08\iMpc$ because non-linear contributions to the measured matter--matter field become important, so we use a smooth theory calculation from \camb \citep{Lewis2000}. In practice this is not a problem at these smaller scales because the measurement is not noise dominated, and in any case the noise would not benefit from cosmic-variance cancellation since it is not Gaussian. We then also divide by the product of the linear halo bias factors, which we take from fitting a linear bias model to the $k<0.08\iMpc$ portion of the measurement of $P_\mathrm{hh}(M_1, M_2, k)/P^\mathrm{lin}_\mathrm{mm}(k)$. We also investigated taking the linear bias from the \cite{Tinker2010} fitting function, and found similar results\footnote{In principle, one could also use a linear bias emulator \citep[\eg][]{Valcin2019, McClintock2019b}.}. However, we noticed discrepancies if we take the linear bias from a less accurate source, for example the peak-background split model of \cite{Tinker2010}.

% /Users/Mead/Physics/Multidark/plotting/BNL_plot.p
\begin{figure*}
\begin{center}
\hspace*{-0.7cm}\includegraphics[height=18.5cm,angle=270]{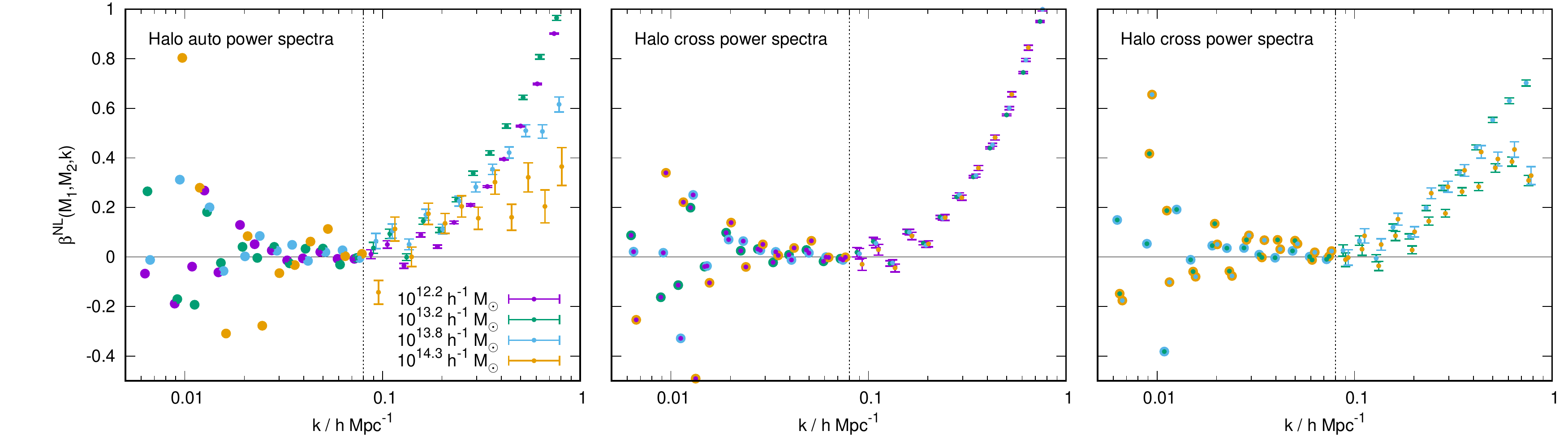}
\end{center}
\caption{Non-linear halo bias function $\Bnl(M_1, M_2, k)$ measured from the \multidark simulation at $z=0$. We show this function measured for auto- and cross-combinations of $4$ halo-mass bins centred at $10^{12.2}$, $10^{13.2}$, $10^{13.8}$ and $10^{14.3}\Msun$. The left-hand panel shows the $4$ auto spectra, while the other panels show the $6$ possible cross spectra, with each point/error set coloured according to the bin colours denoted in the left-hand panel. The dashed vertical line $0.08\iMpc$ indicates our split between those modes we take to be linear and those we take to be non-linear. Error bars, shown only for $k>0.08\iMpc$, show error-on-the-mean power in each $k$ bin in the halo--halo power spectra measured from the simulation, the errors are not shown for $k<0.08\iMpc$ where we enjoy some cosmic variance cancellation because here we divide by the measured linear spectrum. At low $k$, we see that $\Bnl$ tends to zero, as per our expectation. We see non-zero structure start to emerge in $\Bnl$ for $k\gtsim0.08\iMpc$. For $k\sim0.8\iMpc$ we see the function approaches unity for some halo-mass bins, which indicates that this will provide order-unity corrections to analytical calculations around these scales. That the function turns over and starts to decrease at small scales for spectra that involve the higher halo masses is probably because of halo exclusion, which limits how close two haloes can physically be, thus killing the power below scales that correspond to the sum of the two virial radii.}
\label{fig:bnl_1d}
\end{figure*}

% Explanation of one-dimensional non-linear bias figure
In Fig.~\ref{fig:bnl_1d} we show the $\Bnl$ function measured from \multidark for $4$ different mass bins at $z=0$. The measurement is noisy at large scales, but appears to asymptote to zero. Structure in $\Bnl$ becomes visible for $k>0.08\iMpc$ and we observe a prominent, positive detection of the function for $k>0.1\iMpc$ with an amplitude that is dependent on the mass bin. At smaller scales still, the function seems to decay, particularly for the higher mass bins, which we suspect may be because the spatially exclusivity of haloes ensures that the correlation function must be $-1$ at scales smaller than the sum of the virial radii of the two halo populations being correlated. This halo-exclusion condition for small scales in the correlation function will translate in to a small-scale depletion in power for wavenumbers greater than that corresponding to the sum of the virial radii. The two highest-mass bins shown in Fig.~\ref{fig:bnl_1d} correspond to $r_\mathrm{v}\simeq 0.94$ and $1.26\Mpc$ respectively, which correspond to the dips visible in the right-hand panel to within factors of $\sim\pi$.

% ./bin/HMx 68
% load 'plotting/bnl_integrand.p'
\begin{figure*}
\begin{center}
\includegraphics[height=17.5cm,angle=270]{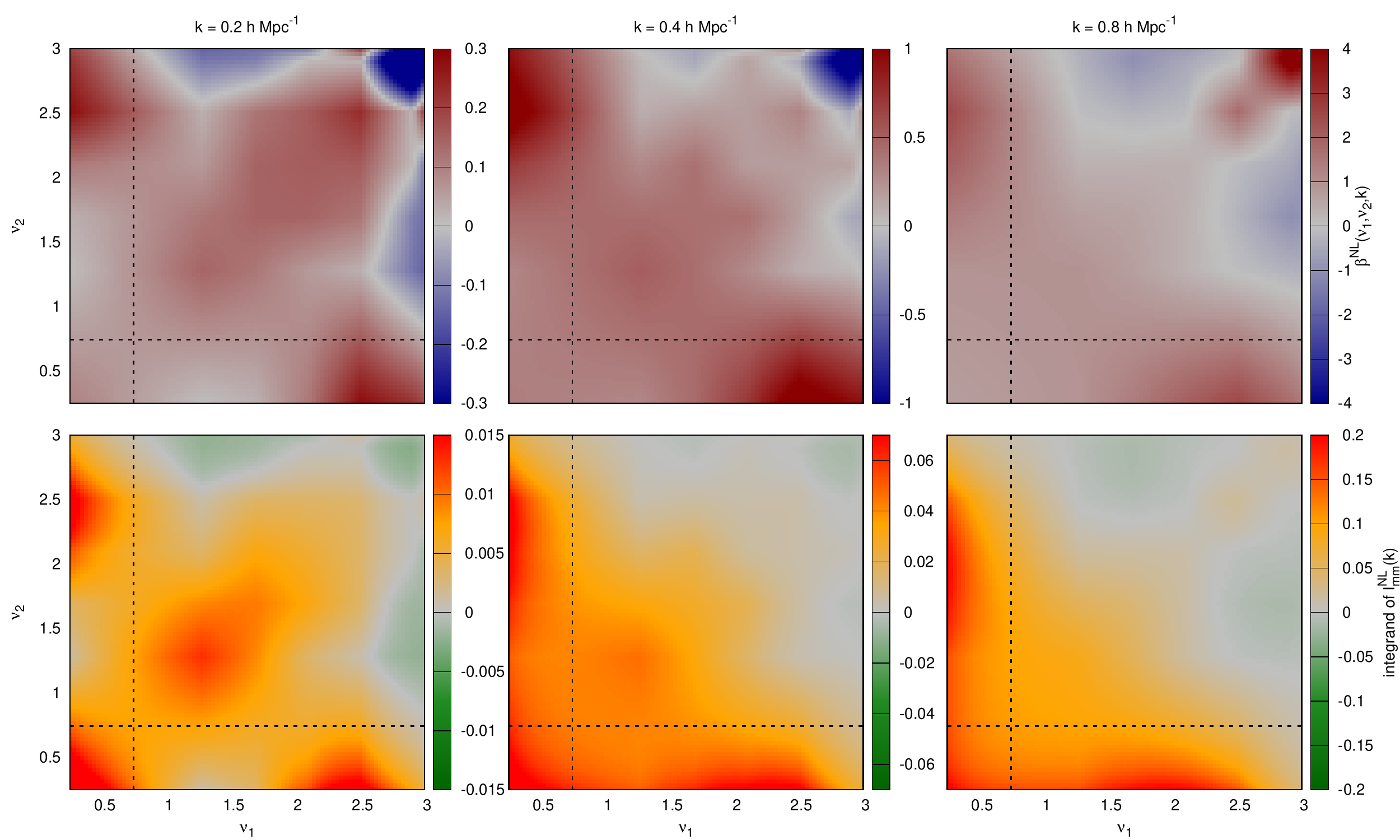}
\end{center}
\caption{The upper set of three panels show the function $\Bnl(\nu_1,\nu_2,k)$ at $k=0.2$ (left), $0.4$, and $0.8\iMpc$ (right), all at $z=0$. The lower three panels show the integrand for the function $I^\mathrm{NL}_\mathrm{mm}(k)$, defined in equation~(\ref{eq:Inl}). This integrand is $\Bnl(\nu_1,\nu_2,k)U_\mathrm{m}(\nu_1,k)U_\mathrm{m}(\nu_2,k)b(\nu_1)b(\nu_2)g(\nu_1)g(\nu_2)$ and it is shown here for the case of the matter--matter power spectrum; for most of the masses and scales shown here $U_\mathrm{m}(\nu,k)\simeq 1$. $\Bnl$ is measured directly from the \multidark simulation in $\nu$ bins and we linearly interpolate between these bins to acquire a continuous function, which leaves some square residual patterns, most visible in the top-left panel. The lowest mass haloes that we measure correspond to $\nu\simeq0.75$, which corresponds to the dashed vertical and horizontal lines in each panel. To get values for $\Bnl$ outside this limit we linearly extrapolate from our measurements at higher halo masses. For reference, for this cosmology at this redshift $\nu=0.5$, $1$, $2$, and $3$ correspond to halo masses: $\log_{10}(M/\Msun)\simeq 10.4$, $12.5$, $14.2$, and $14.9$.}
\label{fig:bnl}
\end{figure*}

% Discussion of the two-dimensional BNL figure
In the top row of Fig.~\ref{fig:bnl} we show $\Bnl$ at $z=0$ as a function of two mass variables as measured from \multidark at $k=0.1$, $0.3$ and $0.8\iMpc$ separately. We parametrize the function in terms of the `peak height', $\nu$, rather than $M$ (equation \ref{eq:nu}) because this pertains directly to our numerical implementation (Section~\ref{sec:results}). For reference, $\nu=0.5$, $1$, $2$ and $3$ correspond to $\simeq10^{10.4}$, $10^{12.5}$, $10^{14.2}$, and $10^{14.9}\Msun$ haloes for this cosmology at $z=0$. We see that, in general, the function increases in amplitude as $k$ increases, as can also be seen in Fig.~\ref{fig:bnl_1d}. We also see the expected reflection symmetry about the $\nu_1=\nu_2$ line. We measured this function from \multidark for $\nu\gtsim0.75$, and for lower values of $\nu$ we linearly extrapolate to get an estimate of $\Bnl$ (we discuss the validity of this later). We are not able to measure $\Bnl$ for $\nu\ltsim0.75$ because this corresponds to haloes below our $50$ particle limit. Note that we only have $25$ bins in $k$ and $8$ bins in $\nu$ for the measurement, and the rest of $\Bnl$ is calculated either via linear interpolation or extrapolation in three dimensions. The locations of high signal in Fig.~\ref{fig:bnl} indicate cross spectra that have particularly strong non-linear halo biases. We note particular intensity in the cross between very low $\nu\sim0.4$ and high $\nu\sim2.5$ haloes. This plausibly originates from low-mass haloes falling on to higher-mass haloes. We also see a negative signal between $\nu\sim2$ and $\nu\sim3$ haloes, which could be a result of high-mass haloes having formed through major mergers leaving a deficit of the haloes that needed to have merged.

\section{Results}
\label{sec:results}

\subsection{Calculation detail}
\label{sec:results_calculation}

% Halo-model ingredient choices
The main results of this paper are halo-model calculations for the matter--matter, matter--halo and halo--halo power spectra when the two-halo term  is calculated using equation~(\ref{eq:new_two_halo_term}). From equation~(\ref{eq:Inl}) we see that this requires us to evaluate a double integral over the $\Bnl(M_1,M_2,k)$ function weighted by halo bias, mass function, and profile functions. Difficulty arises in the numerical integration when one of the fields is `matter' because of substantial contributions from low-mass haloes, well below the mass typically resolved by \nbody simulations, we discuss this in detail in Appendix~\ref{app:numerical_calculations}. In practice, we evaluate this integral by converting $M$ to the dimensionless `peak height', defined in equation~(\ref{eq:nu}). We then use the mass functions $g(\nu)$, normalised such that the integral over all $\nu$ gives unity, which is related to $n(M)$ via
\begin{equation}
g(\nu)\,\mathrm{d}\nu=\frac{M}{\bar\rho}n(M)\,\mathrm{d}M\ .
\label{eq:mass_function}
\end{equation}
For our halo-model calculations we take the form of $g(\nu)$ and $b(\nu)$ from \cite{Tinker2010} and we take the overdensity threshold, $\Delta_\mathrm{v}(z)$, to be that used for halo identification in the \multidark simulations (taken from \citealt{Bryan1998}: $\simeq360\bar\rho$ at $z=0$, asymptoting to $\simeq178\bar\rho$ at high $z$). The mass function and halo bias are defined in such a way that equations~(\ref{eq:mass_normalisation}) and~(\ref{eq:bias_normalisation}) are automatically satisfied. For any power spectrum involving the matter field we must also choose a halo profile, and we adopt the profile of \cite*{Navarro1997}, 
\begin{equation}
\rho(M,r)\propto\frac{1}{r/r_\mathrm{s}(1+r/r_\mathrm{s})^2}\ ,
\label{eq:CDM_profile}
\end{equation}
which is truncated at the virial radius defined such that this encloses an average density of $\Delta_\mathrm{v}(z)$ times the mean density:
\begin{equation}
 M=4\pi r_\mathrm{v}^3\Delta_\mathrm{v}(z)\bar\rho\ .
 \end{equation}
 The virial radius is related to the halo-scale radius, $r_\mathrm{s}$ via the mass-dependent concentration parameter $c=r_\mathrm{v}/r_\mathrm{s}$, which we take from \cite{Duffy2008} and use the appropriate redshift-dependent relation for their full sample of haloes identified using a virial criterion. We are primarily interested in intermediate `quasi-linear' scales and neither the adopted halo profile nor specific concentration-mass relation are important for our results\footnote{We get essentially identical results for everything presented in this paper if we replace the NFW profile with an isothermal, $\propto 1/r^2$, profile.}, which only start to have a significant impact for $k\gtsim1\iMpc$. 
 
 % Lower row of Fig. 2
 The lower row of Fig.~\ref{fig:bnl} shows the $I^\mathrm{NL}_\mathrm{mm}$ integrand, defined in equation~(\ref{eq:Inl}), for the special case of the matter--matter power spectrum. For the scales and mass ranges shown, $W_\mathrm{m}(M, k)\simeq M/\bar\rho$ and therefore the most significant change when going from $\Bnl$ to the integrand for $I^\mathrm{NL}_\mathrm{mm}$ is the suppression at high $\nu$ caused by the halo-mass function. Fig.~\ref{fig:bnl} therefore shows the halo-mass ranges that give additional contributions in our two-halo term from the non-linear halo bias.
 
% Details of evaluating the non-linear bias correction
When evaluating the non-linear bias correction in equation~(\ref{eq:Inl}) we force the correction to be zero for $k<0.08\iMpc$. This is consistent with our earlier choices because we measure our linear halo bias using these scales, and so we are making an implicit assumption that scale-dependent halo bias should be zero here. This choice makes only a small difference to our results because the correction, even when evaluated, is very small for $k<0.08\iMpc$. However, because $\Bnl$ is noise dominated at these scales, if we do not force the correction to zero we transfer this noise into our halo-model calculation.
 
% Two-halo figure
% ./bin/HMx 124
% load 'plotting/bnl_twohalo.p'
\begin{figure}
\begin{center}
\hspace*{-0.30cm}\includegraphics[height=8.5cm,angle=270]{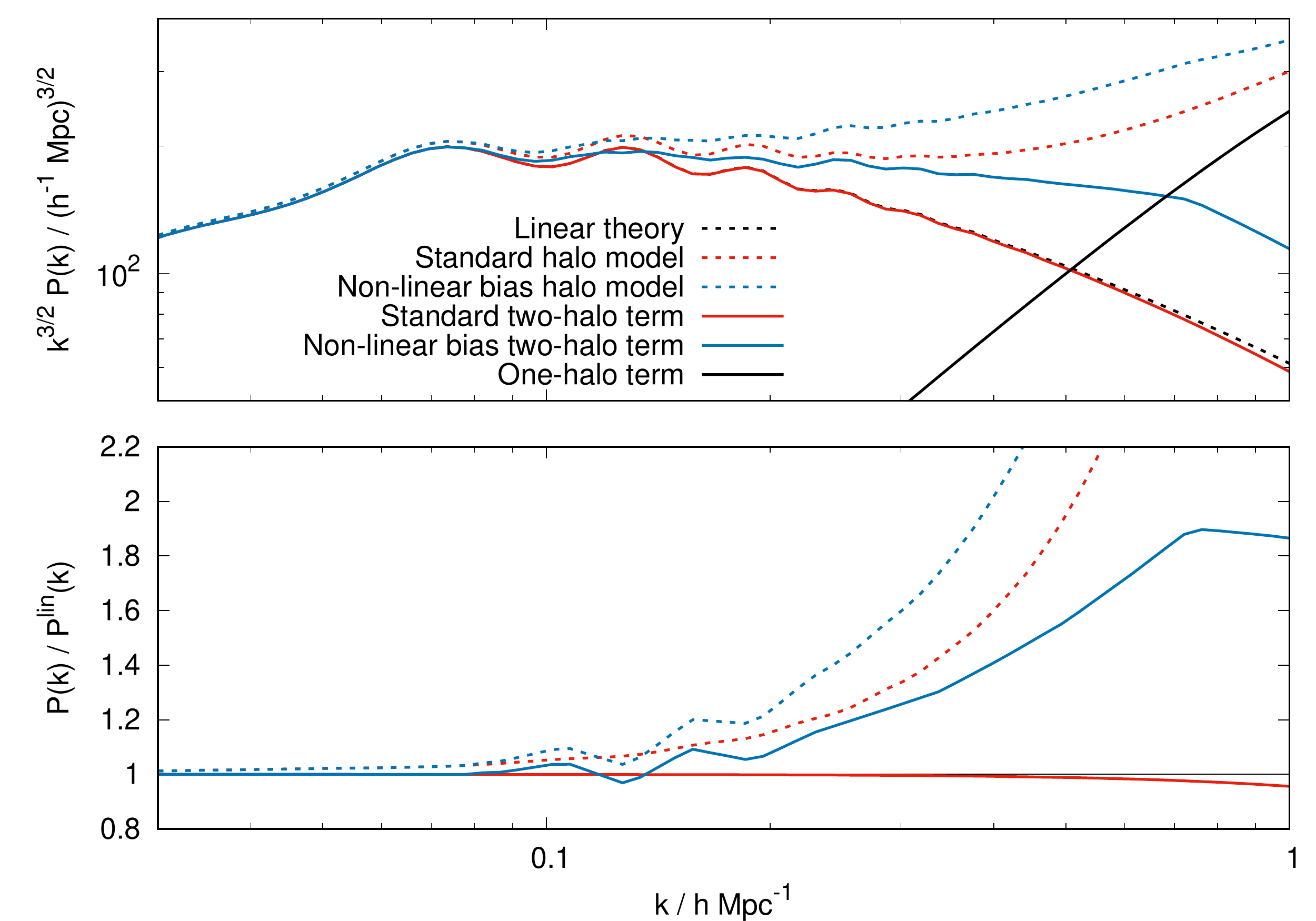}
\end{center}
\caption{Halo model matter--matter power spectrum predictions (dashed) decomposed in to two-halo (solid) and one-halo terms (solid black) for the standard halo model (red) and our improved non-linear halo bias model (blue) at $z=0$. The one-halo term is identical for each model. The linear power (dashed black) is also shown. The top panel shows the power spectra, while the lower panel shows the ratio of the predictions to linear theory. We see that the power in our new two-halo term is approximately double that of the standard at $k\simeq0.7\iMpc$.}
\label{fig:two_halo}
\end{figure}
 
% Discussion of two-halo term figure
In Fig.~\ref{fig:two_halo} we compare the halo-model predictions for the matter--matter power spectrum decomposed in to the different two- and one-halo terms. Scale-dependence compared to linear theory in the standard two-halo term can only ever be a suppression, which arises only due to the scale dependence in the halo window profiles ($W_n(M,k)$ in equation~\ref{eq:standard_two_halo_term}). This small suppression can be seen at the smallest scales shown in Fig.~\ref{fig:two_halo}. In contrast, the inclusion of non-linear halo bias invokes a strong scale dependence in the two-halo term, which boosts the power compared to the standard, starting at $k\sim0.1\iMpc$, and this boost can reach a factor of $\sim2$ at $k\sim0.7\iMpc$. However, such a strong effect is not seen in full in the total power prediction because the maximum of this occurs on scales where the one-halo term (which is identical in each model) starts to dominate the power spectrum prediction. Despite this, the total power in our new model is still boosted by $\sim50$ per cent in the transition region between the two terms compared to the standard model. That power in our two-halo term starts to decay for $k\gtsim0.7\iMpc$ is a combination of halo exclusion effects that are included in our measurement of $\Bnl$, window profiles truncation from equation~(\ref{eq:Inl}), and the fact that we only measure $\Bnl$ for $k\ltsim 0.8\iMpc$ and extrapolate it beyond this. In any case, at such small scales the one-halo term dominates the overall power spectrum.

\subsection{Halo-model power spectra}
\label{sec:results_main}

% Massive plot
% ./bin/HMx 69
% load 'plotting/Multidark_halo_power.p' snap=85
\begin{figure*}
\begin{center}
\includegraphics[height=17.5cm,angle=270]{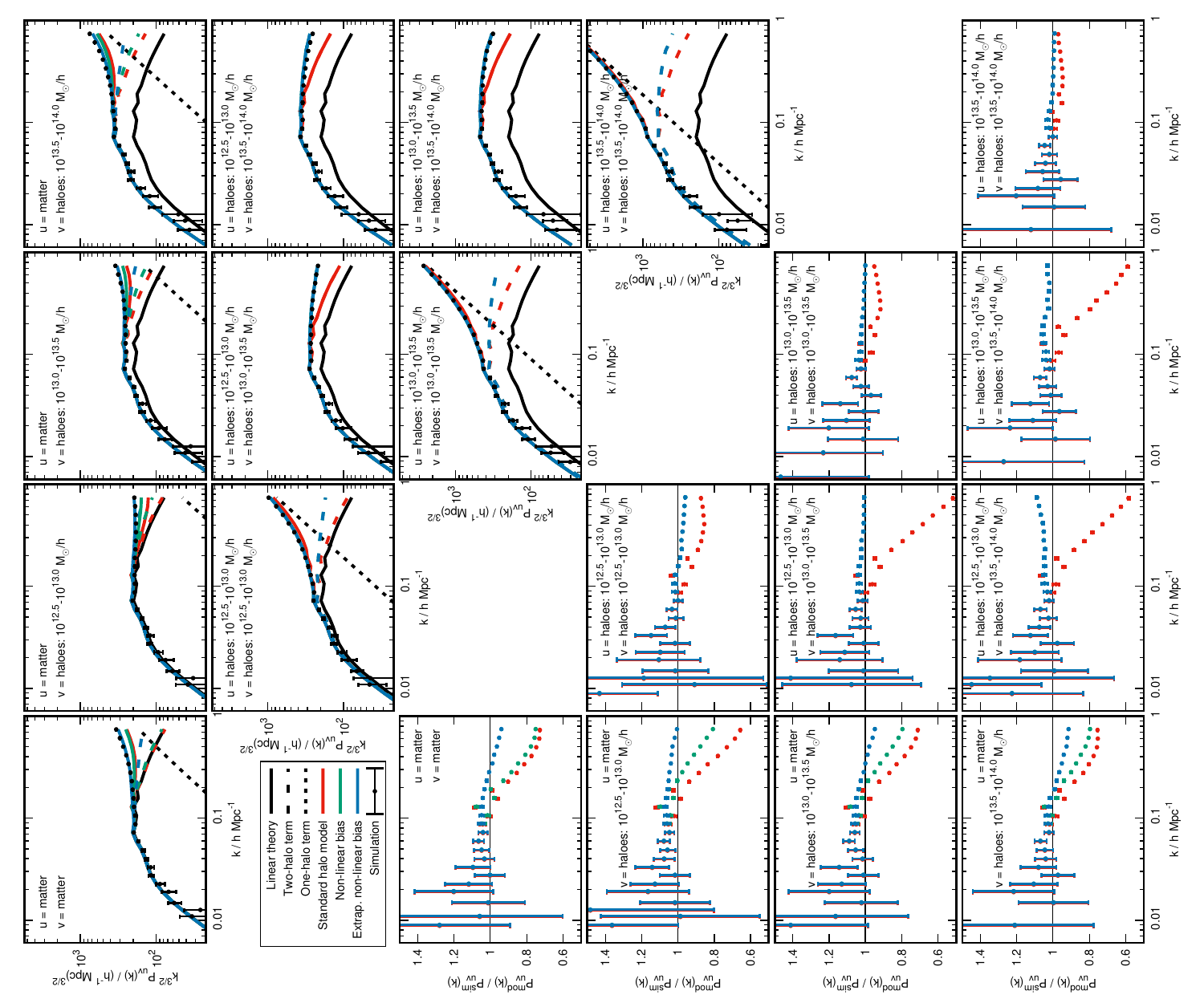}
\end{center}
\caption{Upper-triangular set shows cross-power spectra computed between different cosmological fields using a standard halo-model calculation (red) and our improved method (blue and green) compared to measurements from simulations (black points with errors). The cosmological fields we show are matter overdensity and three sequential halo-mass bins of halo overdensity. The two- (long-dashed coloured) and one-halo (short-dashed black) terms are also shown, but note that the one-halo term is identical for both models. The linear theory matter power spectrum is also shown (solid black). The effect of the improvement to the calculation that is discussed in this paper is prominent in the transition region between the two- and the one-halo terms. This can be better appreciated in the residual panels in the lower-triangular set where the model is divided by the simulated data. For each spectrum we see a clear improvement when using the new halo-model calculation with non-linear halo bias included. Error bars are errors-on-the-mean taken from power spectrum measurements from the simulations.}
\label{fig:Multidark_snap85_halo_power}
\end{figure*}

% Discussion of big multi-panel figure
Fig.~\ref{fig:Multidark_snap85_halo_power} is the main result of this paper, and shows halo-model calculations with and without including $\Bnl$. We show power for auto and cross combinations for three halo mass bins: $10^{12.5}$--$10^{13.0}$, $10^{13.0}$--$10^{13.5}$, $10^{13.5}$--$10^{14.0}\Msun$ and the matter field, all taken from \multidark. Diagonal panels show the auto spectra while off-diagonal panels show the cross spectra. Error bars show error-on-the-mean power in each $k$ bin.

% Discussion of upper triangle set of raw power measurements
The top triangle set show a comparison of the raw power spectrum measurement from simulations to the model predictions. This set of plots is not very useful for investigating the fine details of the performance of our method, but does demonstrate that we are generating predictions that are broadly realistic when compared to the simulated data. Note that in each panel the one-halo term is the same for both the standard and non-linear bias model calculations. The one-halo term is only present in some panels: those where there is an overlap between the haloes in the two fields that make up the power spectrum.

% Discussion of halo-halo power in big figure
The lower triangle set show the ratio of our model predictions to the simulation measurements, with the error bar translated from the simulations into this new space. The standard halo-model prediction is shown together with that from two versions of our calculation that include $\Bnl$. First we discuss the halo--halo auto spectra: in this case we see a small improvement when non-linear halo bias is included in the calculation, but an improvement that corrects the $\sim 10$ per cent low residual that would otherwise be present. If we instead consider the cross spectra between the halo-mass bins, we see a more dramatic improvement, where under predictions in power of $\sim 40$ per cent are almost perfectly cured when one includes the non-linear halo bias. This dramatic difference between the auto- and cross-spectra is due to the halo shot noise, which appears in the auto spectra only and is accounted for via the one-halo term of the standard halo model calculation. The fact that this is absent in the cross spectra between different halo-mass bins allows us to see the true deficit in two-halo power when compared to the linear-bias--linear-power assumption in the standard halo model, which is obscured in the auto spectra by the relatively powerful one-halo term at smaller scales. In some ways, the success of our model for the halo auto- and cross-spectra is not much of a success, given that we essentially use the difference between the standard halo model prediction and the simulations to inform the correction in the non-linear halo bias model. What saves these plots from complete triviality is the fact that the mass bins used are different from those used in our measurement of $\Bnl$ and also that the model predictions come via the full halo-model apparatus, including the choice of \cite{Tinker2010} for the mass function and bias. If there were any serious discrepancies between these choices and reality, these would manifest in the Figure. In this sense, the halo--halo panels of Fig.~\ref{fig:Multidark_snap85_halo_power} provide a useful sanity check and inform us that our halo-model implementation is performing as expected. That the most serious discrepancies occur in the auto spectra is a result of our theoretical mass function not agreeing perfectly with the halo population seen in the simulations, which leads to a slightly different amplitude of the one-halo term in each case. This discrepancy indicates that more accurate halo mass function predictions\footnote{In future one could use the mass-function emulators of either \cite{McClintock2019a} or \cite{Bocquet2020}.} would be useful in further application of this work.

% Discussion of matter--x power in big figure
The more interesting panels of Fig.~\ref{fig:Multidark_snap85_halo_power} are those that show the matter--matter and matter--halo power. These demonstrate the utility of our method, given that the non-linear halo bias correction to the matter originates from an integral over all halo masses. We show two versions of this calculation, one where we restrict the limits of the integral in equation~(\ref{eq:Inl}) to be only over halo masses that we have actually measured from \multidark. The upper limit of \multidark is $\nu\simeq4$ which is effectively infinite from the point of view of the calculation as the results are unchanged if we extrapolate $\Bnl$ above this or fix it to zero. The lower limit of \multidark is $\nu\simeq0.75$ and our results \emph{do} change depending on if we either extrapolate below this limit or fix $\Bnl$ to zero, and we show the impact of these two choices in Fig.~\ref{fig:Multidark_snap85_halo_power}. Note that this choice has no impact on the halo--halo spectra that we show, since all of these correspond to the correction evaluated only in sub-squares of $\Bnl$ and $I^\mathrm{NL}$ shown in Fig.~\ref{fig:bnl} that have been measured well. The halo--matter spectra instead correspond to slices and the matter--matter spectra corresponds to the whole plane. In all cases where matter forms part of a two-point function we see a dramatic improvement in the accuracy of the calculations when the non-linear halo bias is included, particularly when we allow ourselves to extrapolate the measured $\Bnl$ function to all values of $\nu$. In this case the problem of an under-prediction in power in the transition region that plagues the standard halo-model calculation is dramatically ameliorated -- this is the main result of this paper.  In all cases, there are only very small corrections predicted by the model for $0.08<k/\iMpc\ltsim0.1$, which is a result of linear theory and constant bias being a very good approximation for these scales. There is perhaps some small improvement, even for $k\sim0.1\iMpc$, which could arise from $\Bnl$ including some `perturbative' type corrections, \eg pre virialisation, but the data are quite noisy and we are unwilling to draw any firm conclusions about this. For $k>0.1\iMpc$ the correction has a significant impact on our predictions, and this demonstrates that the lack of a proper incorporation of non-linear halo bias in the standard halo model is mostly responsible for the under-prediction of power in the transition region.

% Halo-mass ranges
From Fig.~\ref{fig:Multidark_snap85_halo_power} we note that the required correction in any power spectra that involves haloes is larger for the lower mass halo bins (both in halo--halo and halo--matter spectra). The relatively good performance of halo models that pertain to higher mass haloes must therefore be due to two effects: First, that non-linear halo biasing seems to be intrinsically less important for high-mass haloes, and second that the one-halo term is relatively larger due to the increased shot-noise amplitude arising from the low number density of rare objects, and this one-halo term then obscures the two-halo term on scales where non-linear halo biasing is important. This suggests that traditional halo-model approaches will be more successful when describing the power spectra of fields dominated by higher-mass haloes, and this is broadly the trend seen in the literature. For example, the power spectrum of the tSZ effect has a well understood halo-model interpretation \citep[\eg][]{Komatsu2002, Refregier2002, Battaglia2012b, Horowitz2017}. In the case of tSZ the dominance of the one-halo term is even more extreme due to the extra mass weighting that the electron pressure brings for high-mass haloes \citep[\eg][]{Mead2020}. On the other hand, we would expect a poorer performance of the standard halo model when considering fields that arise from lower-mass haloes. For example, \cite{Addison2013} note that including non-linear halo bias is necessary when modelling the auto spectrum of the cosmic infrared background (CIB), which would make sense given the $\sim10^{12.5}\Msun$ peak halo mass for star formation efficiency \citep[\eg][]{Viero2013, Planck2013XXX, Maniyar2021}. Neutral hydrogen (HI) is known to trace relatively low-mass haloes due to its destruction by energetic feedback processes in massive galaxies, which would suggest that traditional halo model approaches \citep[\eg][]{Padmanabhan2017b} may be less successful for modelling the HI distribution.

% Extrapolation plot
% ./bin/HMx 69
% load 'plotting/Multidark_halo_extrap.p'
\begin{figure}
\begin{center}
\hspace*{0cm}\includegraphics[height=8.5cm, angle=270]{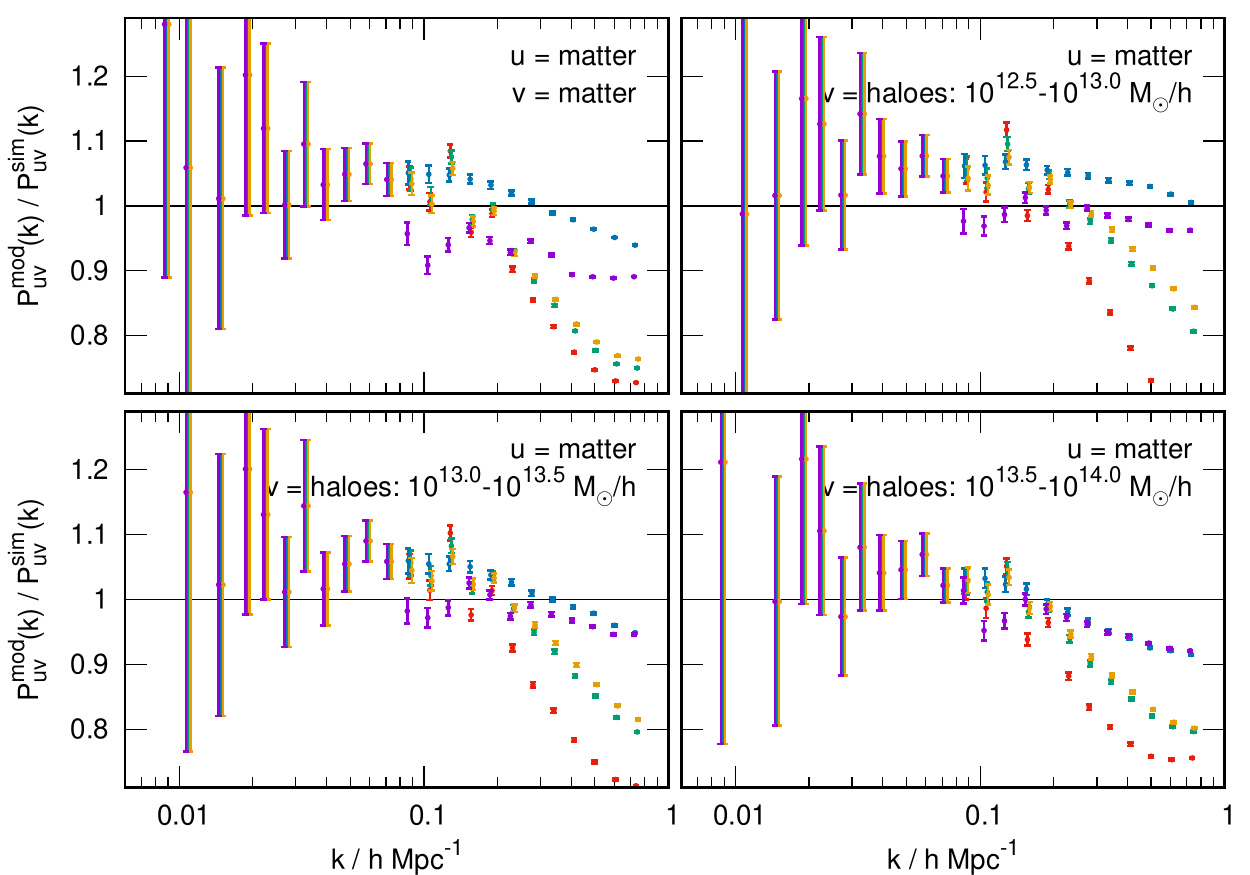}
\end{center}
\caption{Residual power spectra from halo models compared to measurements from simulations at $z=0$ for the matter--matter power spectrum (top left) and matter--halo power spectra for three different halo-mass bins (other panels). In each case we show the standard halo model (red) and then the non-linear halo bias model with $\Bnl$ taken from: \multidark (green); this extrapolated to low halo mass (blue); the combination of \multidark and \bolshoi (yellow); this extrapolated to low halo mass (purple). The difference between green and yellow therefore shows the difference when including the $\sim0.5<\nu<0.75$ haloes from \bolshoi. The difference between purple and blue gives some indication of the robustness of our extrapolation. These panels correspond to the left-hand column of the lower triangle of Fig.~\ref{fig:Multidark_snap85_halo_power}}
\label{fig:extrap}
\end{figure}

\subsection{Extrapolation}
\label{sec:results_extrapolation}

% Effect of extrapolation
We note that a significant fraction of the improvement that we see in the transition region for spectra that involve matter arises only when we allow ourselves to extrapolate our measured $\Bnl$ to lower halo masses: $\nu\ltsim0.75$. The exact fraction of the improvement that depends on these low halo masses varies, but in all of our matter--halo power spectra it is at least half and in our matter--matter it is a factor of $\sim5$. Taken at face value, this would suggest that it is the non-linear bias of low-mass haloes relative to high-mass haloes, and the non-linear bias of low-mass haloes with themselves, that are primarily responsible for the power deficit in the transition region. This agrees somewhat with conclusions drawn by \cite{vanDaalen2015} who demonstrated that significant power arises when matter that is `just outside' the halo virial radius is accounted for. It it plausible that this matter is comprised of infalling, low-mass haloes, and that it is the non-linear biasing (in our language) of these objects that is adding considerable power. Of course, we should also be wary of trusting our extrapolation, particularly given that we are inferring the non-linear bias of roughly half of the mass from the other half, and there may well be complexities regarding the distribution of low-mass haloes that are not captured by linear extrapolation. To partially address this we included data from the \bolshoi simulation in our $\Bnl$ measurement\footnote{Note that here we have used the BDM halo catalogues from \multidark and \bolshoi because \bolshoi has no public \rockstar catalogue.}. \bolshoi is similar to \multidark, but is a smaller $250\Mpc$ cube, so the mass-resolution is better by a factor of $64$, which allows us to measure the non-linear bias of haloes down to $\nu\sim0.5$, albeit with increased noise at any given scale due to the smaller box size. If we do this we get broadly consistent results with those shown in Fig.~\ref{fig:Multidark_snap85_halo_power} with the amount of extrapolation lessened, this is shown in Fig.~\ref{fig:extrap}. The difference between the extrapolation in the two cases is around $5$ per cent in power. The degraded performance that we see for $k\sim0.1\iMpc$ is due to noise from (the small volume of) \bolshoi transferring in to our measurements. This gives us hope that our extrapolation is moderately robust. In any case, even if one is reluctant to trust our extrapolation, the fact that we see any improvement at all in the transition region tells us that non-linear halo bias is at least part of the puzzle surrounding missing power in this region, and does not disprove the hypothesis that it may be entirely responsible.

\subsection{Relation to perturbation theory}
\label{sec:results_perturbation}

% Schemes to include a 'correct' quasi-linear description
The model presented in this paper does not explicitly include beyond-linear perturbation theory, and instead takes a pragmatic approach by measuring the required non-linear halo bias correction directly from \nbody simulation data. We include here a discussion of how our results relate to beyond-linear perturbation theory (see \eg \citealt{Bernardeau2002} and references therein), which is known to be an excellent description of the matter clustering on large scales. For example, one-loop standard perturbation theory (SPT) is sub-per-cent level accurate for standard \LCDM spectra for $k\ltsim0.08\iMpc$ whereas linear theory provides a $\sim 2$ per cent over prediction of power on these `pre-virialisataion' scales. Higher-order perturbative schemes and effective field theories have the potential to push this accuracy to smaller scales \citep[for recent incarnations see \eg][]{Foreman2016a, Seljak2015, Philcox2020}. Taking a schematic approach, assume that one believes that a good large-scale model of the quasi-linear matter--matter power is given by
\begin{equation}
P^\mathrm{QL}_\mathrm{mm}(k)\simeq P^\mathrm{lin}_\mathrm{mm}(k)+P^\mathrm{cor}_\mathrm{mm}(k)\ ,
\end{equation}
where the `cor' term on the right-hand side contains all corrections to linear theory. One could explicitly include this in our approach in two ways: First, the large-scale limit of $\Bnl$ could be changed to enforce $I^\mathrm{NL}_\mathrm{mm}(k\to0)=P^\mathrm{cor}_\mathrm{mm}(k)/P^\mathrm{lin}_\mathrm{mm}(k)$ (in the previous discussion this limit was assumed to be zero). Second, one could replace the factors of $P^\mathrm{lin}_\mathrm{mm}(k)$ that appear on the right-hand side of equation~(\ref{eq:new_two_halo_term}) with $P^\mathrm{QL}_\mathrm{mm}(k)$, such that all of the halo biasing is defined relative to a spectrum that one believes to be more correct at smaller scales. This would entail a redefinition of $\Bnl$, again with the linear power replaced by the quasi-linear power (in equation~\ref{eq:Bnl_rewrite}), and would need the limit $I^\mathrm{NL}_\mathrm{mm}(k\to0)=0$ to be enforced on scales where $P^\mathrm{QL}_\mathrm{mm}(k)$ was thought to be a good description of the power. This redefinition may reduce the scale-dependence of our $\Bnl$ function.

% Our attempt with SPT
We attempted to include the above by taking $P_\mathrm{mm}^\mathrm{cor}(k)$ to be given by one-loop SPT and redefining our biasing relative to this. However, the problem we encountered is that while SPT provides a slightly improved prediction of power for $k<0.1\iMpc$ when it does go wrong, it goes wrong quite spectacularly -- grossly overpredicting the non-linear matter--matter power for $k\gtsim0.2\iMpc$. Given this, we found using linear theory to be more convenient because when it does go wrong it does not go very wrong. In addition, the SPT correction on scales where we might notice it is much smaller than the statistical errors from the limited simulation volume, so would not be noticeable in any of our plots.

% Relation to Smith (2007) paper
In the perturbative bias models of \cite{McDonald2006b} or \cite{Smith2007}, one-loop SPT and an Eulerian bias expansion are taken to be good descriptions of the matter and halo clustering on large scales. This has the advantage that the integral constraints on the Eulerian bias coefficients ensure that the SPT result is returned when integrating over all halo masses ($I^\mathrm{NL}_\mathrm{mm}(k)=0$ in our language). The disadvantage is that this model fails when perturbation theory fails, and this failure can be quite extreme compared to the failures endured by linear theory (with linear bias) at the same scale. Still, it may be possible to incorporate the successes of the \cite{Smith2007} approach with our model by making a split in method based on wavenumber.

% Why not to use non-linear theory
One may also consider using the result from a fitting function (\halofit, \hmcode, or an emulator) for the full non-linear matter--matter power in place of the linear spectrum in equations~(\ref{eq:new_two_halo_term}) and (\ref{eq:Bnl_rewrite}). We do not do this as we wish for our model to be a good description of the power of any combination of fields, including matter--matter. Including the full non-linear matter--matter in the two-halo term therefore leads to a recursion that we would prefer to avoid.

\subsection{Galaxy clustering example}
\label{sec:results_galaxies}
 
% HOD realisation figure
% ./Fortran/Multidark_HOD.e from /Users/Mead/Physics/Multidark
% load 'plotting/HOD_residual.p'
\begin{figure}
\begin{center}
\hspace*{-0.30cm}\includegraphics[height=8.5cm,angle=270]{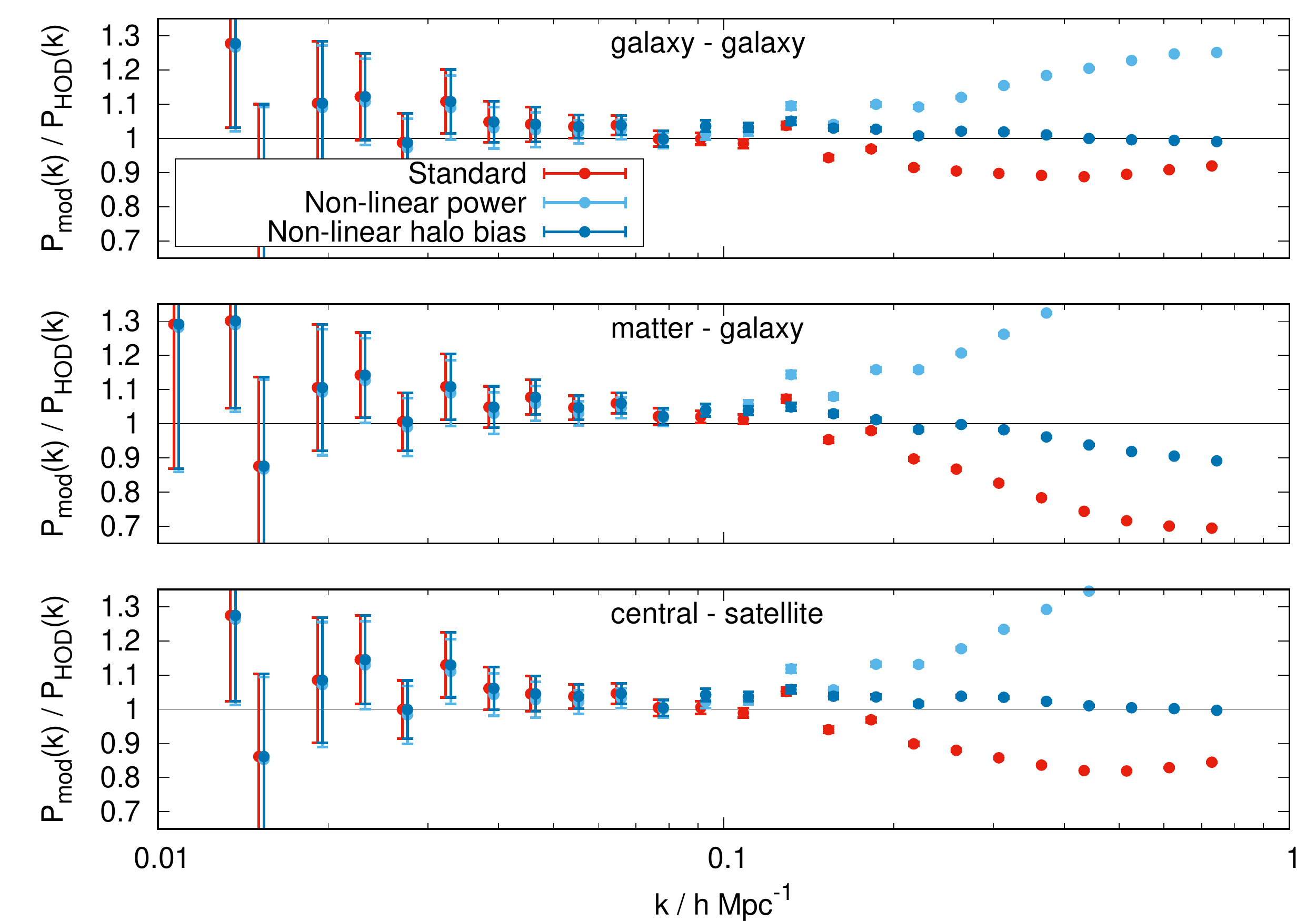}
\end{center}
\caption{Residual power spectra from the standard halo model (red) and our new model that includes non-linear halo bias (dark blue) compared to power spectra from HOD realisations in simulations at $z=0$. We also show power from a halo model where the linear power has been replaced by the non-linear matter--matter power (light blue), thus assuming a linear halo bias with respect to the underlying \emph{non-linear} distribution. The generation of the HOD galaxies is described in the text. We show the galaxy--galaxy (top), galaxy--matter and the central--satellite (bottom) power spectra. Error bars show error-on-the-mean power in each $k$ bin.}
\label{fig:HOD}
\end{figure}

% HOD description
Finally, in Fig.~\ref{fig:HOD}, we demonstrate the importance of our non-linear halo bias correction when comparing analytical halo-model predictions to mock galaxy catalogues that we generate from the \multidark halo catalogues using a simple halo-occupation distribution (HOD) prescription. Our simple HOD catalogues are generated\footnote{We could have used a more complicated HOD, such as \cite{Zheng2005}, but choose to keep the example simple for the sake of pedagogy.} by taking $10^{12.5}$ and $10^{14}\Msun$ as the minimum and maximum halo masses that can host galaxies. Haloes of $10^{12.5}\Msun$ are given a single central galaxy, while haloes above this mass are assigned a number of additional satellite galaxies that depends linearly on the halo mass. For example, a $4\times10^{12.5}\Msun$ halo would be assigned one central and three satellite galaxies. These satellite galaxies are then set to follow an isothermal radial distribution around the halo centre\footnote{Our model is quite insensitive to the exact radial distribution of satellite galaxies on the scales we show in Fig.~\ref{fig:HOD}, instead being primarily sensitive to the occupation number of each halo.}, extending out to the halo virial radius. We make a realisation of this galaxy distribution using the \multidark halo catalogues and we measure the three dimensional galaxy--galaxy, galaxy--matter and central--satellite power spectra. We compare these to predictions from the halo model described in this paper, both with and without the non-linear bias correction in Fig.~\ref{fig:HOD}. Once again, it can be seen that the inclusion of the non-linear halo bias makes a significant improvement to the halo-model predictions around the quasi-linear transition region. We extrapolate $\Bnl$ to low halo masses as described in Section~\ref{sec:results_extrapolation}, but this only affects the galaxy--matter power spectrum because all galaxies live in haloes that have been well measured in $\Bnl$. These results demonstrate the power of the approach advocated in this paper, in that once the $\Bnl$ correction has been measured and characterised, which only needs to be done once using the halo catalogue, any type of HOD prescription can be applied and the novel halo model can be expected to make reasonable predictions. To contrast with this, in Fig.~\ref{fig:HOD} we also show a halo model where we replace the linear power in the standard two-halo term  (equation~\ref{eq:standard_two_halo_term}) with the non-linear matter--matter power (which we take from \halofit of \citealt{Takahashi2012}), thus assuming that the haloes in which the galaxies reside are linearly biased with respect to the underlying \emph{non-linear} matter distribution. This scheme has no physical motivation whatsoever, but is occasionally considered in the literature. We see that this leads to dramatic over-predictions of power for $k>0.1\iMpc$, which is perhaps not surprising given that this model lacks a physical basis and that two one-halo contributions now enter the calculation.

\section{Summary}
\label{sec:summary}

% Introduction
We have proposed an extension to the analytical halo-model formalism that allows for beyond-linear halo bias to be included within an otherwise standard calculation. We have expressed our correction in such a way that the a single new term is added to the (otherwise standard) two-halo term (equations~\ref{eq:new_two_halo_term} and \ref{eq:Inl}), so that it can be easily integrated within existing halo-model implementations. The new correction requires the evaluation of a double integral over halo masses of a new function, which we call $\Bnl$, that accounts for all aspects of the two-point function of haloes that are not accounted for by the simple `linear bias with respect to linear matter field' model. If one prefers to think of halo bias as being defined with respect to the underlying non-linear field then our $\Bnl$ correction can be thought of as an effective bias that incorporates both standard halo biasing and the non-linear evolution of the underlying matter. We have demonstrated that our new corrective term generally boosts power in the transition region between the two- and one-halo terms, which is known to be poorly modelled by traditional halo-model approaches. Our results suggest that most of the power deficit in this region arises from the lack of consideration of beyond-linear halo bias, and that this can be a $\sim 50$ per cent under prediction of power in some cases. We show that when including our correction we get essentially perfect auto spectra for galaxies, and much improved power spectra for matter--galaxy cross spectra. For the matter--matter power spectrum we also get much improved predictions, but we noted a sensitivity in our results to the extrapolation to low-mass haloes that we are required to make of $\Bnl$.

% New paragraph
The advantage of our new approach over traditional approaches is that we capture a genuine, physical effect that is missing from the classic calculation. We also demonstrate that this effect is important, in that it provides corrections of order tens of per cent at scales that are relevant to contemporary surveys. Finally, we show that once the correction has been measured from haloes, it then applies to any two-point function one could calculate via the halo model - it does not need to be recalibrated each time. Of course, the disadvantage is that we must measure a new ingredient from simulations and that currently there exists no fitting form for $\Bnl$.

% Dark Quest
Recently, \cite{Nishimichi2019} have developed the \darkquest emulator, which provides the halo--halo and halo--matter correlation functions with the aim of providing an accurate basis for HOD modelling for galaxy--galaxy and galaxy--matter clustering studies. This has strong similarities to the approach we advocate in this paper, in that the non-linear biasing of haloes is automatically accounted for, and these terms are extracted from \nbody simulations. However, our method is different in some important ways: We retain exactly the standard halo model apparatus on the largest scales, where all power spectra are described by the standard `linear power multiplied by a linear biasing factor'. This allows our non-linear biasing term to appear as a simple addition to an otherwise standard two-halo term, essentially concentrating all new content in the single new $\Bnl$ term. \darkquest, also recovers this large-scale limit, but does so less transparently. We feel that this gives our approach a pedagogical advantage. We also demonstrated that good results can be obtained for spectra that involve the matter field using information from a measured correction that comes \emph{only} from the halo field, which is not considered by \cite{Nishimichi2019}. We note that this was not guaranteed to work, for example the deficit in power in the transition region could have arisen from `one-halo' effects, such as dispersion in halo profiles at fixed mass, or large-scale correlations in shapes. The fact that we see such dramatic improvement tells us that non-linear halo bias is the most important missing ingredient at quasi-linear scales.

% Cosmology and/or redshift dependence
We have not investigated the cosmology dependence of the non-linear halo bias. We suggest that writing the correction in the form given in equation~(\ref{eq:Bnl_rewrite}) may remove some of the cosmology dependence, because it is a ratio of power spectra (see \citealt{Mead2017}) and we also suggest that expressing the correction in terms of peak height, $\nu$, rather than halo mass, may remove additional cosmology dependence. However, we noted that even when expressed in this way our correction retained some redshift dependence (Appendix~\ref{app:additional_redshifts}), which hints at a more complicated relationship with the underlying cosmology. The cosmology dependence could also be investigated using the \darkquest emulator of \cite{Nishimichi2019} and this could even be used to build an emulator for $\Bnl$, but \darkquest is not currently public. One could also use a set of simulations such as the \textsc{quijote} of \cite{Villaescusa-Navarro2020} to build an emulator for $\Bnl$ from scratch. We strongly recommend building such an emulator as a way forward for cosmology, which will allow the halo model to continue to be useful as we enter the epoch of precision measurements. Note carefully that this emulator only needs to be build using information about \emph{halo} clustering, and then the method outlined in this paper can be used to apply the correction to any halo model two-point function.

% Higher-order statistics
One may also consider the application of our work to halo-model descriptions of higher-order statistics of cosmological fields. In this case, we note that $\Bnl$ originates from a two-point function and therefore that beyond-two-point information about halo clustering is absent. Therefore, to successfully describe the $n$-point correlation of a field would require a measurement of the $n$-point version of $\Bnl$. This is in contrast to the standard halo model, where in principle the $n$-point functions all follow from the class halo-model ansatz. However, at large scales it is probable that a perturbative model for non-linear bias could be used, and from this one could consistently build up the $n$-point equivalents of $\Bnl$ at large scales. This could then be augmented with simulation measurements at smaller scales.

% Relation to perturbation theory etc.
Our work was motivated by a consideration of the way halo models are traditionally used to infer properties of halo occupation via cross correlation, be it from tracers or via some emissivity (\eg galaxy or CMB lensing; tSZ; galaxy clustering; CIB), but we note that the error bars on current measurements are quite large \citep[\eg][]{Hill2014, Hojjati2017, Dvornik2018, Tanimura2019b}. This contrasts with more precise work using spectroscopic galaxy surveys, particularly in redshift space, where careful scale cuts can be made. In order to assess the importance of our correction in cases with more exacting accuracy requirements would require us to measure $\Bnl$ more accurately, for example, by using an ensemble of simulations. These simulations could be quite low resolution since all that is required for $\Bnl$ is the halo clustering; our method requires no knowledge of the internal structure of haloes. Once $\Bnl$ has been measured from the halo population it can then be used in any subsequent halo-model calculation. In principle, one could attempt to calculate $\Bnl$ from perturbation theory, but we instead measure the function from the \multidark simulations. We note that $\Bnl$ provides additional power between $0.1\ltsim k/\iMpc \ltsim 1$, thus casting doubt on whether it could ever be described fully perturbatively. However, we suggest that in future a perturbative description at low $k$ could be combined with measurements from simulations at smaller scales. In our calculations, we evaluate $\Bnl$ from an interpolator that is constructed from our measurements, but in principle one could also create a fitting function or an emulator for $\Bnl$, in the same spirit that \cite{McClintock2019b} have provided for the linear halo bias and \cite{Valcin2019} have provided for the large-scale non-linear halo bias.

% Warning
We end this paper with a warning: We have demonstrated that the standard halo model systematically underestimates power in the transition region between the two- and one-halo terms, and that there is compelling evidence that \emph{all} the missing power in this region originates from the beyond-linear biasing of haloes. The amount of power underestimated depends on the particular two-point combination under consideration, but seems to be higher for spectra that include lower-mass haloes, and those for which the one-halo term is subdued (\eg cross spectra between fields that arise from different haloes), because a powerful one-halo term has the potential to obscure the effect of non-linear halo biasing. In the specific three-dimensional power spectra investigated in this paper the amount of missing power varied between $10$ and $50$ per cent. Despite this, the standard halo model is often used to draw conclusions from cosmological data sets about the connection between an observable (\eg galaxies, tSZ, CIB) and the host haloes and even about cosmological parameters. These halo models can often be quite complicated, with a plethora of parameters that govern the distribution (be it of galaxies or some emissivity) within the halo. It is entirely possible that conclusions drawn from such models will be significantly biased if they rely on data that is modelled by the transition region of a standard halo model calculation. This will be particularly true if there are `physical' terms in these models that can add power in the transition region, which is often the case when dealing with two-dimensional models of projected three-dimensional fields, because one does not then enjoy the $k$ localisation of physical effects as in three dimensions. We therefore recommend that $\Bnl$ (or something like it) be included in future data analyses that use the halo model. At least, if it is not used, we then recommend excising data that are modelled by the transition region between the two- and one-halo terms to avoid biasing results. Clearly the potential constraining power of data that is depleted by this excision will vary between two-point functions, and we suggest that it will be greater for data sets that probe lower-mass haloes. Finally, if the accuracy of our $\Bnl$ measurement is not sufficient for precision analyses then one could at least use our results to gauge the importance of the quasi-linear regime to a specific measurement.

% Data availability
\section*{Data availability statement}

% Standard from MNRAS
The code and some data used in this work are available at \code. The remaining data underlying this article will be shared on request to the corresponding author.

% Acknowledgements
\section*{Acknowledgements}

% AJM and Licia and useful conversations
AJM has received funding from the Horizon 2020 research and innovation programme of the European Union under the Marie Sk\l{}odowska-Curie grant agreement No. 702971 and support from the European Research Council under grant number 647112. LV acknowledges support by European Union's Horizon 2020 research and innovation programme ERC (BePreSySe, grant agreement 725327) and  Spanish MINECO under project PGC2018-098866-B-I00, FEDER, UE. AJM acknowledges useful conversations with Samuel Brieden, Catherine Heymans, Alex Hall, John Peacock, Tilman Tr\"oster, Oliver Philcox, and David Alonso.

% Required for papers that use Multidark
The CosmoSim database used in this paper is a service by the Leibniz-Institute for Astrophysics Potsdam (AIP). The MultiDark database was developed in cooperation with the Spanish MultiDark Consolider Project CSD2009-00064. The authors gratefully acknowledge the Gauss Centre for Supercomputing e.V. (\web{www.gauss-centre.eu}) and the Partnership for Advanced Supercomputing in Europe (PRACE, www.prace-ri.eu) for funding the MultiDark simulation project by providing computing time on the GCS Supercomputer SuperMUC at Leibniz Supercomputing Centre (LRZ, \web{www.lrz.de}). The \multidark Database used in this paper and the web application providing online access to it were constructed as part of the activities of the German Astrophysical Virtual Observatory as result of a collaboration between the Leibniz-Institute for Astrophysics Potsdam (AIP) and the Spanish MultiDark Consolider Project CSD2009-00064. The Bolshoi and MultiDark simulations were run on the NASA's Pleiades supercomputer at the NASA Ames Research Center. The MultiDark-Planck (MDPL) and the BigMD simulation suite have been performed in the Supermuc supercomputer at LRZ using time granted by PRACE.

\label{lastpage}

%All of this stuff is to make the bibliography look better
%Enclose in 'footnote size' to make it smaller font size
\footnotesize{
%JP added these two things to left align it properly
\setlength{\bibhang}{2.0em}
\setlength\labelwidth{0.0em}
\bibliographystyle{mnras}
\bibliography{meadbib}

\begin{thebibliography}{}
\makeatletter
\relax
\def\mn@urlcharsother{\let\do\@makeother \do\$\do\&\do\#\do\^\do\_\do\%\do\~}
\def\mn@doi{\begingroup\mn@urlcharsother \@ifnextchar [ {\mn@doi@}
  {\mn@doi@[]}}
\def\mn@doi@[#1]#2{\def\@tempa{#1}\ifx\@tempa\@empty \href
  {http://dx.doi.org/#2} {doi:#2}\else \href {http://dx.doi.org/#2} {#1}\fi
  \endgroup}
\def\mn@eprint#1#2{\mn@eprint@#1:#2::\@nil}
\def\mn@eprint@arXiv#1{\href {http://arxiv.org/abs/#1} {{\tt arXiv:#1}}}
\def\mn@eprint@dblp#1{\href {http://dblp.uni-trier.de/rec/bibtex/#1.xml}
  {dblp:#1}}
\def\mn@eprint@#1:#2:#3:#4\@nil{\def\@tempa {#1}\def\@tempb {#2}\def\@tempc
  {#3}\ifx \@tempc \@empty \let \@tempc \@tempb \let \@tempb \@tempa \fi \ifx
  \@tempb \@empty \def\@tempb {arXiv}\fi \@ifundefined
  {mn@eprint@\@tempb}{\@tempb:\@tempc}{\expandafter \expandafter \csname
  mn@eprint@\@tempb\endcsname \expandafter{\@tempc}}}

\bibitem[\protect\citeauthoryear{{Addison}, {Dunkley}  \& {Spergel}}{{Addison}
  et~al.}{2012}]{Addison2012}
{Addison} G.~E.,  {Dunkley} J.,   {Spergel} D.~N.,  2012, \mn@doi [\mnras]
  {10.1111/j.1365-2966.2012.21664.x}, \href
  {http://adsabs.harvard.edu/abs/2012MNRAS.427.1741A} {427, 1741}

\bibitem[\protect\citeauthoryear{{Addison}, {Dunkley}  \& {Bond}}{{Addison}
  et~al.}{2013}]{Addison2013}
{Addison} G.~E.,  {Dunkley} J.,   {Bond} J.~R.,  2013, \mn@doi [\mnras]
  {10.1093/mnras/stt1703}, \href
  {http://adsabs.harvard.edu/abs/2013MNRAS.436.1896A} {436, 1896}

\bibitem[\protect\citeauthoryear{{Agarwal}, {Abdalla}, {Feldman}, {Lahav}  \&
  {Thomas}}{{Agarwal} et~al.}{2012}]{Agarwal2012}
{Agarwal} S.,  {Abdalla} F.~B.,  {Feldman} H.~A.,  {Lahav} O.,   {Thomas}
  S.~A.,  2012, \mn@doi [\mnras] {10.1111/j.1365-2966.2012.21326.x}, \href
  {http://adsabs.harvard.edu/abs/2012MNRAS.424.1409A} {424, 1409}

\bibitem[\protect\citeauthoryear{{Agarwal}, {Abdalla}, {Feldman}, {Lahav}  \&
  {Thomas}}{{Agarwal} et~al.}{2014}]{Agarwal2014}
{Agarwal} S.,  {Abdalla} F.~B.,  {Feldman} H.~A.,  {Lahav} O.,   {Thomas}
  S.~A.,  2014, \mn@doi [\mnras] {10.1093/mnras/stu090}, \href
  {http://adsabs.harvard.edu/abs/2014MNRAS.439.2102A} {439, 2102}

\bibitem[\protect\citeauthoryear{{Battaglia}, {Bond}, {Pfrommer}  \&
  {Sievers}}{{Battaglia} et~al.}{2012}]{Battaglia2012b}
{Battaglia} N.,  {Bond} J.~R.,  {Pfrommer} C.,   {Sievers} J.~L.,  2012,
  \mn@doi [\apj] {10.1088/0004-637X/758/2/75}, \href
  {http://adsabs.harvard.edu/abs/2012ApJ...758...75B} {758, 75}

\bibitem[\protect\citeauthoryear{{Behroozi}, {Wechsler}  \& {Wu}}{{Behroozi}
  et~al.}{2013}]{Behroozi2013}
{Behroozi} P.~S.,  {Wechsler} R.~H.,   {Wu} H.-Y.,  2013, \mn@doi [\apj]
  {10.1088/0004-637X/762/2/109}, \href
  {https://ui.adsabs.harvard.edu/abs/2013ApJ...762..109B} {762, 109}

\bibitem[\protect\citeauthoryear{Bernardeau, Colombi, Gazta\~{n}aga  \&
  Scoccimarro}{Bernardeau et~al.}{2002}]{Bernardeau2002}
Bernardeau F.,  Colombi S.,  Gazta\~{n}aga E.,   Scoccimarro R.,  2002, \mn@doi
  [Physics Reports] {10.1016/S0370-1573(02)00135-7}, 367, 1

\bibitem[\protect\citeauthoryear{{Bocquet}, {Heitmann}, {Habib}, {Lawrence},
  {Uram}, {Frontiere}, {Pope}  \& {Finkel}}{{Bocquet}
  et~al.}{2020}]{Bocquet2020}
{Bocquet} S.,  {Heitmann} K.,  {Habib} S.,  {Lawrence} E.,  {Uram} T.,
  {Frontiere} N.,  {Pope} A.,   {Finkel} H.,  2020, \mn@doi [\apj]
  {10.3847/1538-4357/abac5c}, \href
  {https://ui.adsabs.harvard.edu/abs/2020ApJ...901....5B} {901, 5}

\bibitem[\protect\citeauthoryear{Bryan \& Norman}{Bryan \&
  Norman}{1998}]{Bryan1998}
Bryan G.~L.,  Norman M.~L.,  1998, \mn@doi [\apj] {10.1086/305262}, 495, 80

\bibitem[\protect\citeauthoryear{{Cacciato}, {van den Bosch}, {More}, {Li},
  {Mo}  \& {Yang}}{{Cacciato} et~al.}{2009}]{Cacciato2009}
{Cacciato} M.,  {van den Bosch} F.~C.,  {More} S.,  {Li} R.,  {Mo} H.~J.,
  {Yang} X.,  2009, \mn@doi [\mnras] {10.1111/j.1365-2966.2008.14362.x}, \href
  {http://adsabs.harvard.edu/abs/2009MNRAS.394..929C} {394, 929}

\bibitem[\protect\citeauthoryear{{Cacciato}, {Lahav}, {van den Bosch},
  {Hoekstra}  \& {Dekel}}{{Cacciato} et~al.}{2012}]{Cacciato2012}
{Cacciato} M.,  {Lahav} O.,  {van den Bosch} F.~C.,  {Hoekstra} H.,   {Dekel}
  A.,  2012, \mn@doi [\mnras] {10.1111/j.1365-2966.2012.21762.x}, \href
  {http://adsabs.harvard.edu/abs/2012MNRAS.426..566C} {426, 566}

\bibitem[\protect\citeauthoryear{{Chen} \& {Afshordi}}{{Chen} \&
  {Afshordi}}{2020}]{Chen2020}
{Chen} A.~Y.,  {Afshordi} N.,  2020, \mn@doi [\prd]
  {10.1103/PhysRevD.101.103522}, \href
  {https://ui.adsabs.harvard.edu/abs/2020PhRvD.101j3522C} {101, 103522}

\bibitem[\protect\citeauthoryear{Cooray \& Sheth}{Cooray \&
  Sheth}{2002}]{Cooray2002}
Cooray A.,  Sheth R.,  2002, \mn@doi [Physics Reports]
  {10.1016/S0370-1573(02)00276-4}, 372, 1

\bibitem[\protect\citeauthoryear{{Duffy}, {Schaye}, {Kay}  \& {Dalla
  Vecchia}}{{Duffy} et~al.}{2008}]{Duffy2008}
{Duffy} A.~R.,  {Schaye} J.,  {Kay} S.~T.,   {Dalla Vecchia} C.,  2008, \mn@doi
  [\mnras] {10.1111/j.1745-3933.2008.00537.x}, \href
  {http://adsabs.harvard.edu/abs/2008MNRAS.390L..64D} {390, L64}

\bibitem[\protect\citeauthoryear{{Dvornik} et~al.,}{{Dvornik}
  et~al.}{2018}]{Dvornik2018}
{Dvornik} A.,  et~al., 2018, \mn@doi [\mnras] {10.1093/mnras/sty1502}, \href
  {https://ui.adsabs.harvard.edu/abs/2018MNRAS.479.1240D} {479, 1240}

\bibitem[\protect\citeauthoryear{{Fedeli}, {Semboloni}, {Velliscig}, {Van
  Daalen}, {Schaye}  \& {Hoekstra}}{{Fedeli} et~al.}{2014}]{Fedeli2014b}
{Fedeli} C.,  {Semboloni} E.,  {Velliscig} M.,  {Van Daalen} M.,  {Schaye} J.,
   {Hoekstra} H.,  2014, \mn@doi [\jcap] {10.1088/1475-7516/2014/08/028}, \href
  {http://adsabs.harvard.edu/abs/2014JCAP...08..028F} {8, 28}

\bibitem[\protect\citeauthoryear{{Foreman}, {Perrier}  \& {Senatore}}{{Foreman}
  et~al.}{2016}]{Foreman2016a}
{Foreman} S.,  {Perrier} H.,   {Senatore} L.,  2016, \mn@doi [\jcap]
  {10.1088/1475-7516/2016/05/027}, \href
  {https://ui.adsabs.harvard.edu/abs/2016JCAP...05..027F} {2016, 027}

\bibitem[\protect\citeauthoryear{{Garcia}, {Rozo}, {Becker}  \&
  {More}}{{Garcia} et~al.}{2020}]{Garcia2020}
{Garcia} R.,  {Rozo} E.,  {Becker} M.~R.,   {More} S.,  2020, arXiv e-prints,
  \href {https://ui.adsabs.harvard.edu/abs/2020arXiv200612751G} {p.
  arXiv:2006.12751}

\bibitem[\protect\citeauthoryear{{Giocoli}, {Bartelmann}, {Sheth}  \&
  {Cacciato}}{{Giocoli} et~al.}{2010}]{Giocoli2010}
{Giocoli} C.,  {Bartelmann} M.,  {Sheth} R.~K.,   {Cacciato} M.,  2010, \mn@doi
  [\mnras] {10.1111/j.1365-2966.2010.17108.x}, \href
  {http://adsabs.harvard.edu/abs/2010MNRAS.408..300G} {408, 300}

\bibitem[\protect\citeauthoryear{{Hang}, {Alam}, {Peacock}  \& {Cai}}{{Hang}
  et~al.}{2021}]{Hang2021}
{Hang} Q.,  {Alam} S.,  {Peacock} J.~A.,   {Cai} Y.-C.,  2021, \mn@doi [\mnras]
  {10.1093/mnras/staa3738}, \href
  {https://ui.adsabs.harvard.edu/abs/2021MNRAS.501.1481H} {501, 1481}

\bibitem[\protect\citeauthoryear{{Hayashi} \& {White}}{{Hayashi} \&
  {White}}{2008}]{Hayashi2008}
{Hayashi} E.,  {White} S. D.~M.,  2008, \mn@doi [\mnras]
  {10.1111/j.1365-2966.2008.13371.x}, \href
  {https://ui.adsabs.harvard.edu/abs/2008MNRAS.388....2H} {388, 2}

\bibitem[\protect\citeauthoryear{{Hill} \& {Spergel}}{{Hill} \&
  {Spergel}}{2014}]{Hill2014}
{Hill} J.~C.,  {Spergel} D.~N.,  2014, \mn@doi [\jcap]
  {10.1088/1475-7516/2014/02/030}, \href
  {http://adsabs.harvard.edu/abs/2014JCAP...02..030H} {2, 030}

\bibitem[\protect\citeauthoryear{{Hill}, {Baxter}, {Lidz}, {Greco}  \&
  {Jain}}{{Hill} et~al.}{2018}]{Hill2018}
{Hill} J.~C.,  {Baxter} E.~J.,  {Lidz} A.,  {Greco} J.~P.,   {Jain} B.,  2018,
  \mn@doi [\prd] {10.1103/PhysRevD.97.083501}, \href
  {http://adsabs.harvard.edu/abs/2018PhRvD..97h3501H} {97, 083501}

\bibitem[\protect\citeauthoryear{{Hojjati} et~al.,}{{Hojjati}
  et~al.}{2017}]{Hojjati2017}
{Hojjati} A.,  et~al., 2017, \mn@doi [\mnras] {10.1093/mnras/stx1659}, \href
  {http://adsabs.harvard.edu/abs/2017MNRAS.471.1565H} {471, 1565}

\bibitem[\protect\citeauthoryear{{Horowitz} \& {Seljak}}{{Horowitz} \&
  {Seljak}}{2017}]{Horowitz2017}
{Horowitz} B.,  {Seljak} U.,  2017, \mn@doi [\mnras] {10.1093/mnras/stx766},
  \href {http://adsabs.harvard.edu/abs/2017MNRAS.469..394H} {469, 394}

\bibitem[\protect\citeauthoryear{{Klypin} \& {Holtzman}}{{Klypin} \&
  {Holtzman}}{1997}]{Klypin1997}
{Klypin} A.,  {Holtzman} J.,  1997, ArXiv Astrophysics e-prints, \href
  {http://adsabs.harvard.edu/abs/1997astro.ph.12217K} {}

\bibitem[\protect\citeauthoryear{{Klypin}, {Trujillo-Gomez}  \&
  {Primack}}{{Klypin} et~al.}{2011}]{Klypin2011}
{Klypin} A.~A.,  {Trujillo-Gomez} S.,   {Primack} J.,  2011, \mn@doi [\apj]
  {10.1088/0004-637X/740/2/102}, \href
  {http://adsabs.harvard.edu/abs/2011ApJ...740..102K} {740, 102}

\bibitem[\protect\citeauthoryear{{Knabenhans} et~al.,}{{Knabenhans}
  et~al.}{2019}]{Knabenhans2019}
{Knabenhans} M.,  et~al., 2019, \mn@doi [\mnras] {10.1093/mnras/stz197}, \href
  {https://ui.adsabs.harvard.edu/abs/2019MNRAS.484.5509K} {484, 5509}

\bibitem[\protect\citeauthoryear{{Knebe} et~al.,}{{Knebe}
  et~al.}{2011}]{Knebe2011}
{Knebe} A.,  et~al., 2011, \mn@doi [\mnras] {10.1111/j.1365-2966.2011.18858.x},
  \href {http://adsabs.harvard.edu/abs/2011MNRAS.415.2293K} {415, 2293}

\bibitem[\protect\citeauthoryear{{Komatsu} \& {Seljak}}{{Komatsu} \&
  {Seljak}}{2002}]{Komatsu2002}
{Komatsu} E.,  {Seljak} U.,  2002, \mn@doi [\mnras]
  {10.1046/j.1365-8711.2002.05889.x}, \href
  {http://adsabs.harvard.edu/abs/2002MNRAS.336.1256K} {336, 1256}

\bibitem[\protect\citeauthoryear{{Koukoufilippas}, {Alonso}, {Bilicki}  \&
  {Peacock}}{{Koukoufilippas} et~al.}{2020}]{Koukoufilippas2020}
{Koukoufilippas} N.,  {Alonso} D.,  {Bilicki} M.,   {Peacock} J.~A.,  2020,
  \mn@doi [\mnras] {10.1093/mnras/stz3351}, \href
  {https://ui.adsabs.harvard.edu/abs/2020MNRAS.491.5464K} {491, 5464}

\bibitem[\protect\citeauthoryear{{Lawrence}, {Heitmann}, {White}, {Higdon},
  {Wagner}, {Habib}  \& {Williams}}{{Lawrence} et~al.}{2010}]{Lawrence2010}
{Lawrence} E.,  {Heitmann} K.,  {White} M.,  {Higdon} D.,  {Wagner} C.,
  {Habib} S.,   {Williams} B.,  2010, \mn@doi [\apj]
  {10.1088/0004-637X/713/2/1322}, \href
  {http://adsabs.harvard.edu/abs/2010ApJ...713.1322L} {713, 1322}

\bibitem[\protect\citeauthoryear{{Lawrence} et~al.,}{{Lawrence}
  et~al.}{2017}]{Lawrence2017}
{Lawrence} E.,  et~al., 2017, \mn@doi [\apj] {10.3847/1538-4357/aa86a9}, \href
  {http://adsabs.harvard.edu/abs/2017ApJ...847...50L} {847, 50}

\bibitem[\protect\citeauthoryear{Lewis, Challinor  \& Lasenby}{Lewis
  et~al.}{2000}]{Lewis2000}
Lewis A.,  Challinor A.,   Lasenby A.,  2000, \mn@doi [\apj] {10.1086/309179},
  538, 473

\bibitem[\protect\citeauthoryear{{Ma}, {Van Waerbeke}, {Hinshaw}, {Hojjati},
  {Scott}  \& {Zuntz}}{{Ma} et~al.}{2015}]{Ma2015}
{Ma} Y.-Z.,  {Van Waerbeke} L.,  {Hinshaw} G.,  {Hojjati} A.,  {Scott} D.,
  {Zuntz} J.,  2015, \mn@doi [\jcap] {10.1088/1475-7516/2015/09/046}, \href
  {http://adsabs.harvard.edu/abs/2015JCAP...09..046M} {9, 046}

\bibitem[\protect\citeauthoryear{{Mandelbaum}, {Tasitsiomi}, {Seljak},
  {Kravtsov}  \& {Wechsler}}{{Mandelbaum} et~al.}{2005}]{Mandelbaum2005}
{Mandelbaum} R.,  {Tasitsiomi} A.,  {Seljak} U.,  {Kravtsov} A.~V.,
  {Wechsler} R.~H.,  2005, \mn@doi [\mnras] {10.1111/j.1365-2966.2005.09417.x},
  \href {http://adsabs.harvard.edu/abs/2005MNRAS.362.1451M} {362, 1451}

\bibitem[\protect\citeauthoryear{{Maniyar}, {B{\'e}thermin}  \&
  {Lagache}}{{Maniyar} et~al.}{2021}]{Maniyar2021}
{Maniyar} A.,  {B{\'e}thermin} M.,   {Lagache} G.,  2021, \mn@doi [\aap]
  {10.1051/0004-6361/202038790}, \href
  {https://ui.adsabs.harvard.edu/abs/2021A&A...645A..40M} {645, A40}

\bibitem[\protect\citeauthoryear{{McClintock} et~al.,}{{McClintock}
  et~al.}{2019a}]{McClintock2019b}
{McClintock} T.,  et~al., 2019a, arXiv e-prints, \href
  {https://ui.adsabs.harvard.edu/abs/2019arXiv190713167M} {p. arXiv:1907.13167}

\bibitem[\protect\citeauthoryear{{McClintock} et~al.,}{{McClintock}
  et~al.}{2019b}]{McClintock2019a}
{McClintock} T.,  et~al., 2019b, \mn@doi [\apj] {10.3847/1538-4357/aaf568},
  \href {https://ui.adsabs.harvard.edu/abs/2019ApJ...872...53M} {872, 53}

\bibitem[\protect\citeauthoryear{{McDonald}}{{McDonald}}{2006}]{McDonald2006b}
{McDonald} P.,  2006, \mn@doi [\prd] {10.1103/PhysRevD.74.103512}, \href
  {https://ui.adsabs.harvard.edu/abs/2006PhRvD..74j3512M} {74, 103512}

\bibitem[\protect\citeauthoryear{{Mead}}{{Mead}}{2017}]{Mead2017}
{Mead} A.~J.,  2017, \mn@doi [\mnras] {10.1093/mnras/stw2312}, \href
  {http://adsabs.harvard.edu/abs/2017MNRAS.464.1282M} {464, 1282}

\bibitem[\protect\citeauthoryear{{Mead}, {Peacock}, {Heymans}, {Joudaki}  \&
  {Heavens}}{{Mead} et~al.}{2015}]{Mead2015b}
{Mead} A.~J.,  {Peacock} J.~A.,  {Heymans} C.,  {Joudaki} S.,   {Heavens}
  A.~F.,  2015, \mn@doi [\mnras] {10.1093/mnras/stv2036}, \href
  {http://adsabs.harvard.edu/abs/2015MNRAS.454.1958M} {454, 1958}

\bibitem[\protect\citeauthoryear{{Mead}, {Heymans}, {Lombriser}, {Peacock},
  {Steele}  \& {Winther}}{{Mead} et~al.}{2016}]{Mead2016}
{Mead} A.~J.,  {Heymans} C.,  {Lombriser} L.,  {Peacock} J.~A.,  {Steele}
  O.~I.,   {Winther} H.~A.,  2016, \mn@doi [\mnras] {10.1093/mnras/stw681},
  \href {http://adsabs.harvard.edu/abs/2016MNRAS.459.1468M} {459, 1468}

\bibitem[\protect\citeauthoryear{{Mead}, {Tr{\"o}ster}, {Heymans}, {Van
  Waerbeke}  \& {McCarthy}}{{Mead} et~al.}{2020}]{Mead2020}
{Mead} A.~J.,  {Tr{\"o}ster} T.,  {Heymans} C.,  {Van Waerbeke} L.,
  {McCarthy} I.~G.,  2020, \mn@doi [\aap] {10.1051/0004-6361/202038308}, \href
  {https://ui.adsabs.harvard.edu/abs/2020A&A...641A.130M} {641, A130}

\bibitem[\protect\citeauthoryear{{Mead}, {Brieden}, {Tr{\"o}ster}  \&
  {Heymans}}{{Mead} et~al.}{2021}]{Mead2021a}
{Mead} A.~J.,  {Brieden} S.,  {Tr{\"o}ster} T.,   {Heymans} C.,  2021, \mn@doi
  [\mnras] {10.1093/mnras/stab082}, \href
  {https://ui.adsabs.harvard.edu/abs/2021MNRAS.502.1401M} {502, 1401}

\bibitem[\protect\citeauthoryear{{Mohammed} \& {Seljak}}{{Mohammed} \&
  {Seljak}}{2014}]{Mohammed2014a}
{Mohammed} I.,  {Seljak} U.,  2014, \mn@doi [\mnras] {10.1093/mnras/stu1972},
  \href {https://ui.adsabs.harvard.edu/abs/2014MNRAS.445.3382M} {445, 3382}

\bibitem[\protect\citeauthoryear{{Nakamura} \& {Suto}}{{Nakamura} \&
  {Suto}}{1997}]{Nakamura1997}
{Nakamura} T.~T.,  {Suto} Y.,  1997, \mn@doi [Progress of Theoretical Physics]
  {10.1143/PTP.97.49}, \href
  {http://adsabs.harvard.edu/abs/1997PThPh..97...49N} {97, 49}

\bibitem[\protect\citeauthoryear{Navarro, Frenk  \& White}{Navarro
  et~al.}{1997}]{Navarro1997}
Navarro J.~F.,  Frenk C.~S.,   White S. D.~M.,  1997, \mn@doi [\apj]
  {10.1086/304888}, 490, 493

\bibitem[\protect\citeauthoryear{{Nishimichi} et~al.,}{{Nishimichi}
  et~al.}{2019}]{Nishimichi2019}
{Nishimichi} T.,  et~al., 2019, \mn@doi [\apj] {10.3847/1538-4357/ab3719},
  \href {https://ui.adsabs.harvard.edu/abs/2019ApJ...884...29N} {884, 29}

\bibitem[\protect\citeauthoryear{{Padmanabhan}, {Refregier}  \&
  {Amara}}{{Padmanabhan} et~al.}{2017}]{Padmanabhan2017b}
{Padmanabhan} H.,  {Refregier} A.,   {Amara} A.,  2017, \mn@doi [\mnras]
  {10.1093/mnras/stx979}, \href
  {http://adsabs.harvard.edu/abs/2017MNRAS.469.2323P} {469, 2323}

\bibitem[\protect\citeauthoryear{{Peacock} \& {Smith}}{{Peacock} \&
  {Smith}}{2000}]{Peacock2000}
{Peacock} J.~A.,  {Smith} R.~E.,  2000, \mn@doi [\mnras]
  {10.1046/j.1365-8711.2000.03779.x}, \href
  {http://adsabs.harvard.edu/abs/2000MNRAS.318.1144P} {318, 1144}

\bibitem[\protect\citeauthoryear{{Philcox}, {Spergel}  \&
  {Villaescusa-Navarro}}{{Philcox} et~al.}{2020}]{Philcox2020}
{Philcox} O. H.~E.,  {Spergel} D.~N.,   {Villaescusa-Navarro} F.,  2020,
  \mn@doi [\prd] {10.1103/PhysRevD.101.123520}, \href
  {https://ui.adsabs.harvard.edu/abs/2020PhRvD.101l3520P} {101, 123520}

\bibitem[\protect\citeauthoryear{{Planck Collaboration}}{{Planck
  Collaboration}}{2014}]{Planck2013XXX}
{Planck Collaboration} 2014, \mn@doi [\aap] {10.1051/0004-6361/201322093},
  \href {https://ui.adsabs.harvard.edu/abs/2014A&A...571A..30P} {571, A30}

\bibitem[\protect\citeauthoryear{{Prada}, {Klypin}, {Cuesta}, {Betancort-Rijo}
  \& {Primack}}{{Prada} et~al.}{2012}]{Prada2012}
{Prada} F.,  {Klypin} A.~A.,  {Cuesta} A.~J.,  {Betancort-Rijo} J.~E.,
  {Primack} J.,  2012, \mn@doi [\mnras] {10.1111/j.1365-2966.2012.21007.x},
  \href {http://adsabs.harvard.edu/abs/2012MNRAS.423.3018P} {423, 3018}

\bibitem[\protect\citeauthoryear{{Refregier} \& {Teyssier}}{{Refregier} \&
  {Teyssier}}{2002}]{Refregier2002}
{Refregier} A.,  {Teyssier} R.,  2002, \mn@doi [\prd]
  {10.1103/PhysRevD.66.043002}, \href
  {http://adsabs.harvard.edu/abs/2002PhRvD..66d3002R} {66, 043002}

\bibitem[\protect\citeauthoryear{{Riebe} et~al.,}{{Riebe}
  et~al.}{2013}]{Riebe2013}
{Riebe} K.,  et~al., 2013, \mn@doi [Astronomische Nachrichten]
  {10.1002/asna.201211900}, \href
  {http://adsabs.harvard.edu/abs/2013AN....334..691R} {334, 691}

\bibitem[\protect\citeauthoryear{{Schmidt}}{{Schmidt}}{2016}]{Schmidt2016}
{Schmidt} F.,  2016, \mn@doi [\prd] {10.1103/PhysRevD.93.063512}, \href
  {https://ui.adsabs.harvard.edu/abs/2016PhRvD..93f3512S} {93, 063512}

\bibitem[\protect\citeauthoryear{{Seljak}}{{Seljak}}{2000}]{Seljak2000}
{Seljak} U.,  2000, \mn@doi [\mnras] {10.1046/j.1365-8711.2000.03715.x}, \href
  {http://adsabs.harvard.edu/abs/2000MNRAS.318..203S} {318, 203}

\bibitem[\protect\citeauthoryear{{Seljak} \& {Vlah}}{{Seljak} \&
  {Vlah}}{2015}]{Seljak2015}
{Seljak} U.,  {Vlah} Z.,  2015, \mn@doi [\prd] {10.1103/PhysRevD.91.123516},
  \href {http://adsabs.harvard.edu/abs/2015PhRvD..91l3516S} {91, 123516}

\bibitem[\protect\citeauthoryear{Sheth \& Tormen}{Sheth \&
  Tormen}{1999}]{Sheth1999}
Sheth R.~K.,  Tormen G.,  1999, \mn@doi [\mnras]
  {10.1046/j.1365-8711.1999.02692.x}, 308, 119

\bibitem[\protect\citeauthoryear{{Smith} \& {Markovic}}{{Smith} \&
  {Markovic}}{2011}]{Smith2011b}
{Smith} R.~E.,  {Markovic} K.,  2011, \mn@doi [\prd]
  {10.1103/PhysRevD.84.063507}, \href
  {http://adsabs.harvard.edu/abs/2011PhRvD..84f3507S} {84, 063507}

\bibitem[\protect\citeauthoryear{{Smith} \& {Watts}}{{Smith} \&
  {Watts}}{2005}]{Smith2005}
{Smith} R.~E.,  {Watts} P.~I.~R.,  2005, \mn@doi [\mnras]
  {10.1111/j.1365-2966.2005.09053.x}, \href
  {http://adsabs.harvard.edu/abs/2005MNRAS.360..203S} {360, 203}

\bibitem[\protect\citeauthoryear{{Smith} et~al.,}{{Smith}
  et~al.}{2003}]{Smith2003}
{Smith} R.~E.,  et~al., 2003, \mn@doi [\mnras]
  {10.1046/j.1365-8711.2003.06503.x}, \href
  {http://adsabs.harvard.edu/abs/2003MNRAS.341.1311S} {341, 1311}

\bibitem[\protect\citeauthoryear{{Smith}, {Scoccimarro}  \& {Sheth}}{{Smith}
  et~al.}{2007}]{Smith2007}
{Smith} R.~E.,  {Scoccimarro} R.,   {Sheth} R.~K.,  2007, \mn@doi [\prd]
  {10.1103/PhysRevD.75.063512}, \href
  {http://adsabs.harvard.edu/abs/2007PhRvD..75f3512S} {75, 063512}

\bibitem[\protect\citeauthoryear{{Takahashi}, {Sato}, {Nishimichi}, {Taruya}
  \& {Oguri}}{{Takahashi} et~al.}{2012}]{Takahashi2012}
{Takahashi} R.,  {Sato} M.,  {Nishimichi} T.,  {Taruya} A.,   {Oguri} M.,
  2012, \mn@doi [\apj] {10.1088/0004-637X/761/2/152}, \href
  {http://adsabs.harvard.edu/abs/2012ApJ...761..152T} {761, 152}

\bibitem[\protect\citeauthoryear{{Tanimura}, {Aghanim}, {Douspis}, {Beelen}  \&
  {Bonjean}}{{Tanimura} et~al.}{2019}]{Tanimura2019b}
{Tanimura} H.,  {Aghanim} N.,  {Douspis} M.,  {Beelen} A.,   {Bonjean} V.,
  2019, \mn@doi [\aap] {10.1051/0004-6361/201833413}, \href
  {https://ui.adsabs.harvard.edu/abs/2019A&A...625A..67T} {625, A67}

\bibitem[\protect\citeauthoryear{{Tinker}, {Weinberg}, {Zheng}  \&
  {Zehavi}}{{Tinker} et~al.}{2005}]{Tinker2005}
{Tinker} J.~L.,  {Weinberg} D.~H.,  {Zheng} Z.,   {Zehavi} I.,  2005, \mn@doi
  [\apj] {10.1086/432084}, \href
  {http://adsabs.harvard.edu/abs/2005ApJ...631...41T} {631, 41}

\bibitem[\protect\citeauthoryear{{Tinker}, {Robertson}, {Kravtsov}, {Klypin},
  {Warren}, {Yepes}  \& {Gottl{\"o}ber}}{{Tinker} et~al.}{2010}]{Tinker2010}
{Tinker} J.~L.,  {Robertson} B.~E.,  {Kravtsov} A.~V.,  {Klypin} A.,  {Warren}
  M.~S.,  {Yepes} G.,   {Gottl{\"o}ber} S.,  2010, \mn@doi [\apj]
  {10.1088/0004-637X/724/2/878}, \href
  {http://adsabs.harvard.edu/abs/2010ApJ...724..878T} {724, 878}

\bibitem[\protect\citeauthoryear{{Valageas} \& {Nishimichi}}{{Valageas} \&
  {Nishimichi}}{2011}]{Valageas2011}
{Valageas} P.,  {Nishimichi} T.,  2011, \mn@doi [\aap]
  {10.1051/0004-6361/201015685}, \href
  {http://adsabs.harvard.edu/abs/2011A%26A...527A..87V} {527, A87}

\bibitem[\protect\citeauthoryear{{Valcin}, {Villaescusa-Navarro}, {Verde}  \&
  {Raccanelli}}{{Valcin} et~al.}{2019}]{Valcin2019}
{Valcin} D.,  {Villaescusa-Navarro} F.,  {Verde} L.,   {Raccanelli} A.,  2019,
  \mn@doi [\jcap] {10.1088/1475-7516/2019/12/057}, \href
  {https://ui.adsabs.harvard.edu/abs/2019JCAP...12..057V} {2019, 057}

\bibitem[\protect\citeauthoryear{{Viero} et~al.,}{{Viero}
  et~al.}{2013}]{Viero2013}
{Viero} M.~P.,  et~al., 2013, \mn@doi [\apj] {10.1088/0004-637X/779/1/32},
  \href {https://ui.adsabs.harvard.edu/abs/2013ApJ...779...32V} {779, 32}

\bibitem[\protect\citeauthoryear{{Villaescusa-Navarro}
  et~al.,}{{Villaescusa-Navarro} et~al.}{2020}]{Villaescusa-Navarro2020}
{Villaescusa-Navarro} F.,  et~al., 2020, \mn@doi [\apjs]
  {10.3847/1538-4365/ab9d82}, \href
  {https://ui.adsabs.harvard.edu/abs/2020ApJS..250....2V} {250, 2}

\bibitem[\protect\citeauthoryear{{Voivodic}, {Rubira}  \& {Lima}}{{Voivodic}
  et~al.}{2020}]{Voivodic2020}
{Voivodic} R.,  {Rubira} H.,   {Lima} M.,  2020, \mn@doi [\jcap]
  {10.1088/1475-7516/2020/10/033}, \href
  {https://ui.adsabs.harvard.edu/abs/2020JCAP...10..033V} {2020, 033}

\bibitem[\protect\citeauthoryear{{Wolz}, {Murray}, {Blake}  \& {Wyithe}}{{Wolz}
  et~al.}{2019}]{Wolz2019}
{Wolz} L.,  {Murray} S.~G.,  {Blake} C.,   {Wyithe} J.~S.,  2019, \mn@doi
  [\mnras] {10.1093/mnras/sty3142}, \href
  {https://ui.adsabs.harvard.edu/abs/2019MNRAS.484.1007W} {484, 1007}

\bibitem[\protect\citeauthoryear{{Zheng} et~al.,}{{Zheng}
  et~al.}{2005}]{Zheng2005}
{Zheng} Z.,  et~al., 2005, \mn@doi [\apj] {10.1086/466510}, \href
  {http://adsabs.harvard.edu/abs/2005ApJ...633..791Z} {633, 791}

\bibitem[\protect\citeauthoryear{{van Daalen} \& {Schaye}}{{van Daalen} \&
  {Schaye}}{2015}]{vanDaalen2015}
{van Daalen} M.~P.,  {Schaye} J.,  2015, \mn@doi [\mnras]
  {10.1093/mnras/stv1456}, \href
  {http://adsabs.harvard.edu/abs/2015MNRAS.452.2247V} {452, 2247}

\bibitem[\protect\citeauthoryear{{van den Bosch}, {More}, {Cacciato}, {Mo}  \&
  {Yang}}{{van den Bosch} et~al.}{2013}]{vandenBosch2013}
{van den Bosch} F.~C.,  {More} S.,  {Cacciato} M.,  {Mo} H.,   {Yang} X.,
  2013, \mn@doi [\mnras] {10.1093/mnras/sts006}, \href
  {http://adsabs.harvard.edu/abs/2013MNRAS.430..725V} {430, 725}

\makeatother
\end{thebibliography}
}

\normalsize
\appendix

\section{Numerical calculations}
\label{app:numerical_calculations}

% Raison d'etre
In this paper we have demonstrated that the accuracy of analytical halo-model calculations can be substantially improved by including the non-linear bias of haloes in the two-halo term. Unfortunately, this increases the computational complexity of a calculation and also provides some additional numerical hurdles to overcome in comparison with the standard calculation. In this Appendix we present some of the technical details of how we evaluate our new non-linear bias term.

% Standard two-halo problem
We first remind the reader of a well-known numerical problem (and a solution) that is encountered when trying to evaluate the standard two-halo term for matter--field combinations:
\begin{equation}
P^{2\mathrm{H}}_{uv}(k)=P^\mathrm{lin}_\mathrm{mm}(k)
\prod_{n=u,v}\left[\int_0^\infty W_n(M,k)b(M)n(M)\,\mathrm{d}M\right]\ .
\label{eq:standard_two_halo_term_appendix}
\end{equation}
In principle, one ought to integrate over all halo masses but usually the integration range is restricted to be finite so that the properties of very high and low-mass haloes do not need to be specified. The mass function purports to describe all haloes, however, most popular mass functions \citep[\eg][]{Sheth1999, Tinker2010} propose that a  large fraction of mass is locked within very low-mass haloes, if taken literally. The reality is that very little is known about the ultra low-mass halo population at $z=0$, and the inferences about this population that are made by standard mass functions are a result of the condition that all mass reside within haloes,
\begin{equation}
\int_0^\infty Mn(M)\,\mathrm{d}M=\bar\rho\ ,
\label{eq:mass_normalisation_appendix}
\end{equation}
being imposed on the fitting function by hand\footnote{This condition is often, but not always, imposed on functional forms that are fitted to halo mass functions measured from \nbody simulations.}. For a matter field, we know that in the $k\to0$ limit the integral in equation~(\ref{eq:standard_two_halo_term_appendix}) must equal unity, equivalently:
\begin{equation}
\int_0^\infty Mb(M)n(M)\,\mathrm{d}M=\bar\rho\ .
\label{eq:bias_normalisation_appendix}
\end{equation}
but this requires an unreasonably low lower limit on the integral. As discussed in \cite{Schmidt2016} and \cite{Mead2020} one solution is to enforce equation~(\ref{eq:bias_normalisation_appendix}), which can be achieved in practice by assuming that all mass in haloes below the lower integration limit, $M_\mathrm{min}$, is contained in haloes of mass exactly $M_\mathrm{min}$:
\begin{equation}
n(M) \to n(M)\Theta(M-M_\mathrm{min})+\frac{A(M_\mathrm{min})\Dirac{M-M_\mathrm{min}}}{b(M_\mathrm{min})M_\mathrm{min}/\bar\rho}\ ,
\label{eq:mass_function_modified}
\end{equation}
where
\begin{equation}
A(M_\mathrm{min})=1-\frac{1}{\bar\rho}\int_{M_\mathrm{min}}^\infty Mb(M)n(M)\,\mathrm{d}M\ ,
\label{eq:mass_function_correction}
\end{equation}
which can be easily evaluated and should be some number between zero and unity, with $\sim 0.5$ being typical for standard mass functions and standard integration limits. This procedure is valid because the linear halo bias tends to be constant for low-mass haloes and is sufficient as long as the calculation is not sensitive to the profiles ($k$ dependence) of haloes with $M<M_\mathrm{min}$. Note that this problem generally does not affect the one-halo term (equation~\ref{eq:one_halo_term}) because the extra factor of $W$ within the integral ensures that the this term is dominated by comparatively high-mass haloes that will be included within the range of haloes that is typically integrated over. 

% Non-linear bias
Our improved halo-model calculation requires us to evaluate the double integral
\begin{multline}
I^{\mathrm{NL}}_{uv}(k)=\int_0^\infty\int_0^\infty \Bnl(M_1,M_2,k)\times \\
W_u(M_1,k)W_v(M_2,k)b(M_1)b(M_2)n(M_1)n(M_2)\,\mathrm{d}M_1\mathrm{d}M_2\ .
\label{eq:Inl_appendix}
\end{multline}
If either $W_u$ or $W_v$ pertain to matter we run into the same problems as for the standard two-halo term because we would usually evaluate this term over a finite range in mass. Once again, the upper limit is usually not a concern, so here we will treat it as effectively infinite ($M=10^{16}\Msun$ is a sensible upper limit for a standard \LCDM model at $z=0$) and it is the lower limit that is problematic. However, we can use the same trick and replace both $n(M)$ that appear in equation~(\ref{eq:Inl_appendix}) with that in equation~(\ref{eq:mass_function_modified}). This splits equation~(\ref{eq:Inl_appendix}) in to $4$ terms:
\begin{equation}
I^{\mathrm{11}}_{uv}(k)=
A^2(M_\mathrm{min})
\frac{W_u(M_\mathrm{min},k)}{M_\mathrm{min}/\bar\rho}
\frac{W_v(M_\mathrm{min},k)}{M_\mathrm{min}/\bar\rho}\ ,
\end{equation}
\begin{multline}
I^{\mathrm{12}}_{uv}(k)=
A(M_\mathrm{min})\frac{W_u(M_\mathrm{min},k)}{M_\mathrm{min}/\bar\rho}\times \\
\int_{M_\mathrm{min}}^\infty \Bnl(M_\mathrm{min}, M_2, k)W_v(M_2, k)b(M_2)n(M_2)\,\mathrm{d}M_2\ ,
\end{multline}
\begin{multline}
I^{\mathrm{21}}_{uv}(k)=
A(M_\mathrm{min})\frac{W_v(M_\mathrm{min},k)}{M_\mathrm{min}/\bar\rho}\times \\
\int_{M_\mathrm{min}}^\infty \Bnl(M_1, M_\mathrm{min}, k)W_u(M_1, k)b(M_1)n(M_1)\,\mathrm{d}M_1\ ,
\end{multline}
\begin{multline}
I^{\mathrm{22}}_{uv}(k)=\int_{M_\mathrm{min}}^\infty\int_{M_\mathrm{min}}^\infty \Bnl(M_1,M_2,k)\times \\
W_u(M_1,k)W_v(M_2,k)b(M_1)b(M_2)n(M_1)n(M_2)\,\mathrm{d}M_1\mathrm{d}M_2\ ,
\label{eq:I22}
\end{multline}
which can all be evaluated over $M\in[M_\mathrm{min},\infty]$. The final result is then given by $I^{\mathrm{NL}}_{uv}=I^{\mathrm{11}}_{uv}+I^{\mathrm{12}}_{uv}+I^{\mathrm{21}}_{uv}+I^{\mathrm{22}}_{uv}$. The terms other than $I^{\mathrm{22}}_{uv}(k)$ only ever have an impact if signal arises from haloes with $M<M_\mathrm{min}$ -- they are identically zero if $W(M_\mathrm{min},k)=0$ (\eg for galaxies when $M_\mathrm{min}$ is below the occupation threshold). The additional terms are typically only evaluated when `matter' appears in a two-point function because significant matter exists in low-mass haloes. We checked that our calculation is only minimally sensitive to $M_\mathrm{min}$ with results only changing by $\sim2$ per cent if we change our fiducial lower limit from $10^{8}$ to either $10^{6}$ or $10^{10}\Msun$.

%Splitting the integrals up into the portion above and below our limiting mass $M_\mathrm{min}$ splits the $M_1$--$M_2$ plane of the integral into four quadrants:

\section{Additional redshifts}
\label{app:additional_redshifts}

% z = 0.5, 1 figure
\begin{figure*}
\begin{center}
\includegraphics[height=17.5cm, angle=270]{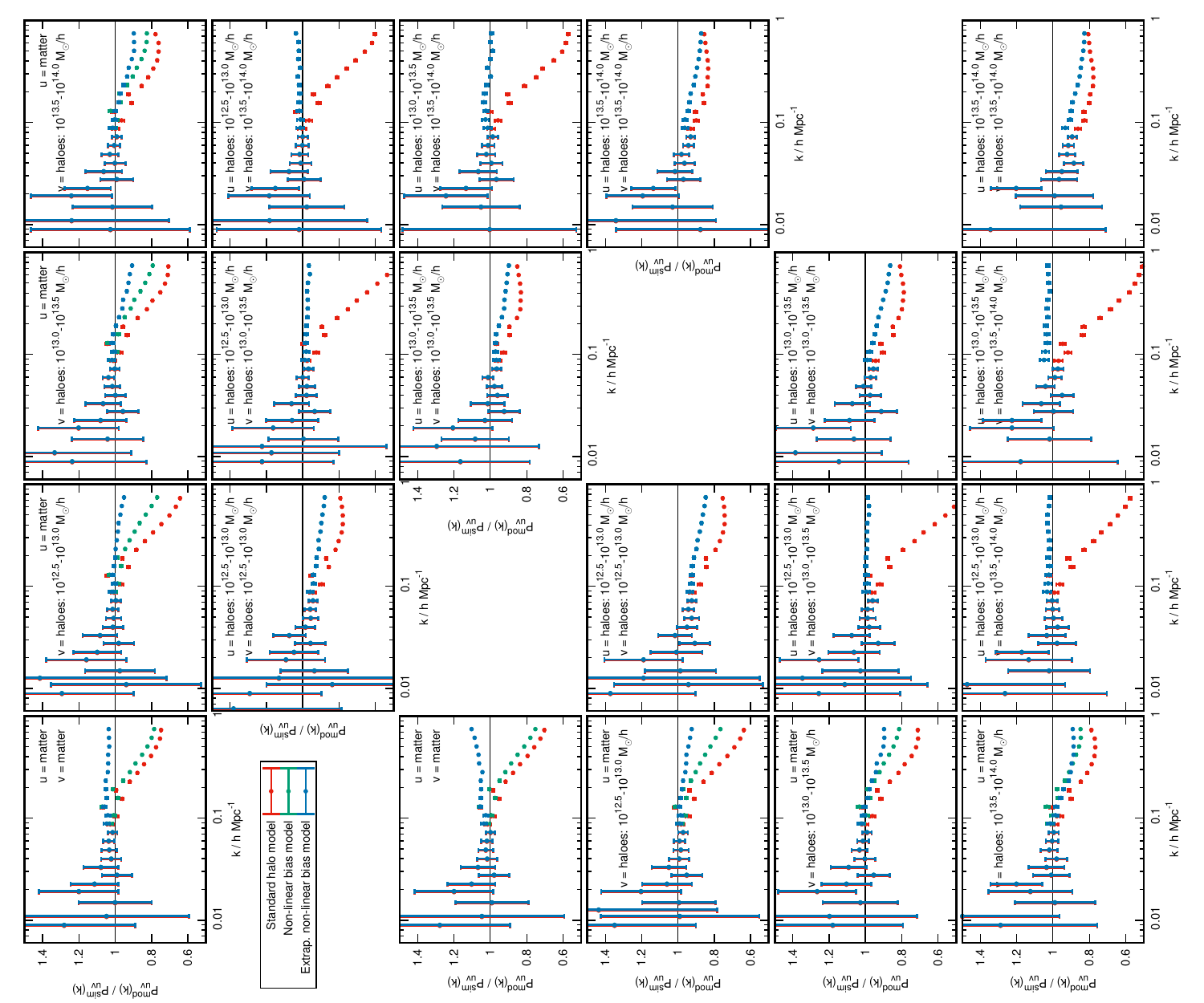}
\end{center}
\caption{As the lower-triangle set of Fig.~\ref{fig:Multidark_snap85_halo_power} but for $z=0.53$ (upper triangle) and $z=1$ (lower triangle). In each case the $\Bnl$ function has been calculated from \multidark data at the redshift in question. Red points with error bars show the standard halo model prediction, while other coloured points show the new model prediction, either extrapolated (blue) or not (green).}
\label{fig:Multidark_halo_residual}
\end{figure*}

% Discuss figure
In Fig.~\ref{fig:Multidark_halo_residual} we show the ratio of halo model predictions for matter--matter, matter--halo and halo--halo power spectra measured from the \multidark simulations. This is the same information as in the lower triangle of Fig.~\ref{fig:Multidark_snap85_halo_power}, but for $z=0.53$ and $z=1$ (upper and lower triangles of Fig.~\ref{fig:Multidark_halo_residual} respectively). We see very similar results to those presented in Fig.~\ref{fig:Multidark_snap85_halo_power} with the accuracy being nearly perfect for the halo--halo cross power (as it should be). The halo auto power show slightly larger discrepancies, but these must arise from the one-halo term, which is telling us that the theoretical mass function must not be a perfect description of the simulations here. For the halo--matter power we see very encouraging results, especially when we allow ourselves to extrapolate the non-linear bias correction to low halo masses. The matter--matter power is also encouraging, although we see a slight tendency for the extrapolated non-linear bias model to overpredict the power in the quasi-linear regime, particularly at $z=1$, which may be indicative of a short coming of our extrapolation.

% Redshift dependence of the correction
For both redshifts shown in Fig.~\ref{fig:Multidark_halo_residual} $\Bnl$ has been re-measured from the simulation data at the redshift in question. We also looked at using the $\Bnl$ function measured at $z=0$ in making predictions at other $z$, but we found that this did not work as well (although the results were still okay), hinting that the correction has a significant redshift dependence. This in turn suggests that $\Bnl$ would have cosmology dependence. This cosmology and redshift dependence is despite the fact that we have chosen to parameterise the correction as a function of $\nu$ (rather than $M$, or $\log M$) and that we have defined $\Bnl$ as in equation~(\ref{eq:Bnl_rewrite}) where we might have hoped that the division by $P_\mathrm{lin}$ would null the cosmology and redshift dependence. We leave any further investigation of this to the future.

\section{Exact linear halo biasing}
\label{app:exact_biasing}

% /Users/Mead/Physics/HODmod/HODmod
% /Users/Mead/Physics/HODmod/plotting/power.p
\begin{figure}
\begin{center}
\hspace*{-0.4cm}\includegraphics[height=9cm, angle=270]{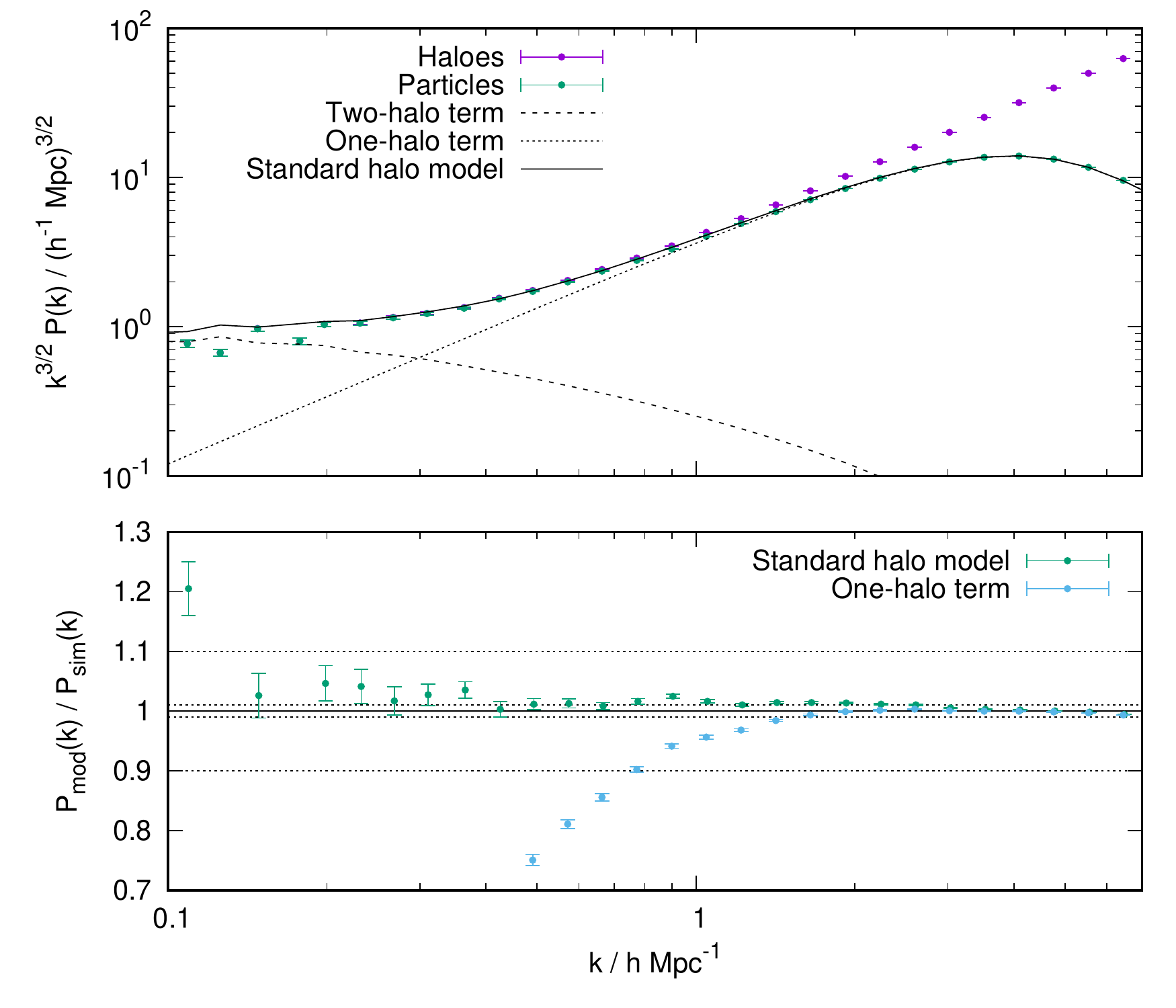}
\end{center}
\caption{Points with error bars in the upper panel show the average particle--particle (green) and halo--halo (purple) power spectra measured from $25$ realisations of our mock universe, which is described in the text. Error bars show error-on-the-mean power in each $k$ bin. The black lines shows a halo model calculation for the particle--particle power (solid), and this broken down into two-halo (dashed) and one-halo (dotted) terms. The lower panel shows the ratio of the model to the simulation for the particle--particle data, with the error bars transferred from the top panel. Here we see that the standard halo model provides an almost perfect description of the data from the (absurd) mock universes.}
\label{fig:HODmod}
\end{figure}

In this Appendix we present a contrived scenario in which the standard halo-model calculation (equations~\ref{eq:standard_two_halo_term} and \ref{eq:one_halo_term}; neglecting non-linear halo bias) is exactly correct. We consider this example to have great pedagogical value, despite its absurdity.

We create a mock universe by first creating $64^3$ `haloes', which we distribute uniform randomly\footnote{Note that this means that haloes can overlap.} within a $100\Mpc$ cube. We consider these haloes to make up all of the mass in our `universe', and we give them each an isothermal density profile corresponding to $\Delta_\mathrm{v}=3$, which we fill with $512$ particles per halo, such that each halo is of equal mass. We then make a Gaussian realisation of a linear displacement field, corresponding to $z=19$ for our \wmap5 cosmology, and displace the haloes accordingly, thus giving them some large-scale clustering. In our example, because we work at $z=19$ in a $100\Mpc$ cube with only $64^3$ haloes, the displacements of each halo are very small compared to the mean inter-halo separation. This makes the Zel'dovich approximation almost perfect, and generates a genuinely linearly biased sample of haloes (albeit with $b=1$). We then measure the power spectrum of haloes--haloes and particles--particles and compare these to analytical halo-model calculations. In the case of particles--particles we subtract the particle shot noise, but we do not do this for haloes-haloes since, in this case, the (halo) shot-noise has physical meaning -- the haloes are the field.

The upper panel of Fig.~\ref{fig:HODmod} shows the power spectra averaged over $10$ realisations. On large scales, the halo and particle power spectra agree perfectly, but at smaller scales we see the particle power spectrum turn over relative to the halo spectrum, which we can attribute to the physical extent of the halo profiles. The advantage of making each halo so bloated ($\Delta_\mathrm{v}=3$, compared to the $~200$ standard value) is that we can actually see the turn over in the one-halo term caused by the halo profiles, which would otherwise be at too small a scale. Intriguingly, we see no power deficit in the two- to one-halo transition region in our mock universe. This is confirmed in the lower panel, where we show the ratio of particle--particle model power to simulations, with the error bars transferred from the simulation measurements. We see that, in this case, the standard halo-model calculation is accurate at the few per-cent level for all wavenumbers shown. Note that there is certainly no $\sim30$ per cent under prediction, which is what is typically seen in measurements from \nbody simulations.

The results presented in this Appendix demonstrate that the analytical halo-model calculation is exactly correct when compared to our mock universe, where all of the approximations that go into deriving it are satisfied. Haloes are linearly biased, exactly spherical with known mass and virial radius (which in our example are the same for all haloes, but we could have drawn these from a mass distribution function) with smooth profiles. This illustrates that there is nothing strange going on when the calculation breaks down when compared to realistic simulations (or the realistic Universe) -- it is simply a reflection of the assumptions behind the calculation not being valid. As demonstrated in this paper, the major missing ingredient at intermediate scales between the standard two- and one-halo terms is the non-linearity of the halo bias.

\end{document}